\documentclass[authoryear,final,times,3p, 12pt]{elsarticle}

\usepackage{hyperref,geometry,subfigure}
 \usepackage{graphicx}
\usepackage{amssymb, upgreek, amsmath, lscape, natbib}
\renewcommand{\eqref}[1]{Eq.~(\ref{#1})}
\renewcommand{\vec}[1]{{\bf #1}}

\journal{Journal of Computational Physics}
\begin{document}
\begin{frontmatter}

\title{Higher-Order Compositional Modeling of Three-phase Flow in 3D Fractured Porous Media Using Cross-flow Equilibrium Approach}

\author{Joachim Moortgat}
\ead{jmoortgat@rerinst.org}
\author{Abbas Firoozabadi}
\ead{abbas.firoozabadi@yale.edu}
\address{Reservoir Engineering Research Institute, 595 Lytton Ave, Suite B, Palo Alto, CA 94301, USA.}

\begin{abstract}
Numerical simulation of multiphase compositional flow in fractured porous media, when all the species can transfer between the phases, is a real challenge. Despite the broad applications in hydrocarbon reservoir engineering and hydrology, a compositional numerical simulator for three-phase flow in fractured media has not appeared in the literature, to the best of our knowledge. In this work, we present a three-phase fully compositional simulator for fractured media, based on higher-order finite element methods. To achieve computational efficiency, we invoke the cross-flow equilibrium (CFE) concept between discrete fractures and a small neighborhood in the matrix blocks. We adopt the mixed hybrid finite element (MHFE) method to approximate convective Darcy fluxes and the pressure equation. This approach is the most natural choice for flow in fractured media. The mass balance equations are discretized by the discontinuous Galerkin (DG) method, which is perhaps the most efficient approach to capture physical discontinuities in phase properties at the matrix-fracture interfaces and at phase boundaries.  In this work, we account for gravity and Fickian diffusion. The modeling of capillary effects is discussed in a separate paper. We present the mathematical framework, using the implicit-pressure-explicit-composition (IMPEC) scheme, which facilitates rigorous thermodynamic stability analyses and the computation of phase behavior effects to account for transfer of species between the phases.
 A deceptively simple CFL condition is implemented to improve numerical stability and accuracy. We provide six numerical examples at both small and larger scales and in two and three dimensions, to demonstrate powerful features of the formulation.
\end{abstract}


\end{frontmatter}

%% main text
\section{Introduction}
\label{sec::intro}
Many problems in hydrocarbon reservoir engineering, as well as hydrology, involve the flow of multiple distinct phases in fractured porous media. One important example is gas injection in oil reservoirs that have previously been water flooded. Another example is when gas is injected in an oil reservoir and a third hydrocarbon phase develops with intermediate properties to the gas and oil phases. 
When there is significant species exchange between different phases, there is a need for multi-phase compositional simulators. A reliable determination of the number of phases, phase amounts and phase compositions requires an equation of state (EOS) based phase stability analysis and three-phase-split computations. Thermodynamic stability analysis is essential to guarantee that a phase-split solution corresponds to the lowest Gibbs free energy. For the three-phase-split, in particular, there are generally multiple solutions corresponding to local minima. Fully compositional EOS-based commercial simulators for three hydrocarbon phases are currently not available. 

Compositional simulators commonly use Henry's law or similar correlations to predict the $\mathrm{CO}_{2}$ solubility in water (e.g.~\cite{coatswater}). This is a poor approximation in three-phase flow, where it cannot satisfy thermodynamic equilibrium. The $\mathrm{CO}_{2}$ composition in the aqueous phase has to satisfy equality of fugacities of $\mathrm{CO}_{2}$ in all three phases, which cannot be guaranteed by Henry's law. In the presence of $\mathrm{CO}_{2}$-rich gas and oil phases, water is not necessarily saturated with $\mathrm{CO}_{2}$.
In our work, we have developed a rigorous EOS-based three-phase thermodynamics algorithm. The aqueous phase is modeled by the cubic-plus-association (CPA) EOS, including cross-association between $\mathrm{CO}_{2}$ and water molecules and self-association between water molecules \citep{liCPA}. At low temperatures ($T < 350\ \mathrm{K}$), evaporation of water and the mutual solubility between water and hydrocarbon phases can be neglected. To speed up the phase-split computation for water-gas-oil mixtures, we only consider $\mathrm{CO}_{2}$ solubility in water. The CPA-EOS reduces to the \citet{preos} (PR) EOS for the hydrocarbon phases, when they do not contain water.
The three-phase compositional model and details of the phase-split computations were presented in \citet{moortgatRSS} for unfractured domains. Another compositional model for three hydrocarbon phases in unfractured media was considered in \citet{okuno}. 
In this paper, we extend the modeling of both water-oil-gas systems and the flow of three hydrocarbon phases to the considerably more complicated fractured porous media. 

Fractured porous media are challenging because of the large range in spatial scales, permeabilities, fluxes and phase properties. Currently, the most efficient compositional simulators are based on implicit-pressure-explicit-composition (IMPEC) schemes. The explicit mass transport update is constrained by the Courant-Friedrichs-Lewy (CFL) condition on the maximum time-step, which is proportional to the size of grid elements, and inversely proportional to the flux \citep{cfl1}. Fractures generally have small apertures, but may allow large fluxes due to the high fracture permeability. When fractures are discretized the same way as matrix elements, i.e.~single-porosity simulations, the resulting CFL condition is exceedingly small and most problems are numerically intractable. The most commonly used alternatives are dual-porosity or dual-porosity-dual-permeability models, which use two overlapping domains \citep{warrenroot}. 
All the flow is through a sugar-cube configuration of fractures, while the matrix only serves as a storage medium. The flux between fractures and matrix blocks is computed by so-called transfer functions. The dual-porosity and dual-permeability models are adopted in most fractured media studies (for various implementations of varying complexity, see for instance \cite{deswaan, coatsdualsingleporosity,peng, fernando, bennion, cicek, huiyun, ramirez}). The approach is highly efficient, particularly for immiscible and single-phase flow, but suffers from severe limitations for multi-phase compositional flow, when there is significant species exchange by Fickian diffusion between the fractures and the neighboring region in the matrix, gravitational reinfiltration of oil from fractures to matrix blocks, or gravitational and viscous instabilities that may cross fractures. Such complex physical processes may not rigorously be incorporated in transfer functions.

We adopt an alternative approach in which fractures are combined with a small fraction of the matrix blocks on either side in larger computational elements. The assumption is that a large permeability, but small pressure gradient across the fracture-matrix interface results in a transverse flux that instantaneously equilibrates the fracture fluid with the fluid immediately next to it in the matrix. We call this the cross-flow equilibrium approximation, and denote the combined fracture-matrix elements as CF elements. The fluxes across the edges of the CF elements are worked out by integrating the appropriate Darcy fluxes for the matrix and fracture contributions. Variations of the discrete fracture approach were studied analytically by \citet{tan1small, tan2small}, applied to immiscible water injection in fractured media by \citet{karimi2003}, to three-phase black-oil in \citet{geigerfracs, fu} and to single- and two-phase compositional flow in \citet{hoteit2005, hoteit2006b}. 
In those papers, and the examples presented in this paper for three-phase flow, it is demonstrated that the CF approach provides nearly indistinguishable results from fine mesh simulations, but at orders of magnitude lower CPU cost. The CF treatment of fractures allows coarser grids, which translates into large CFL time-steps. The efficiency of the pressure update is also improved, because of the lower contrast in (CF) fracture and matrix permeabilities. In \citet{hoteitcapillary}, the MHFE+DG and CF approach was applied to immiscible two-phase flow in fractured media with capillary pressures. 
In this work, we neglect capillarity for simplicity. The additional complications posed by capillarity in heterogeneous and fractured domains are presented in a separate paper \citep{moortgatcapillarity}.

The paper is organized as follows. In Section~\ref{sec::math} the mathematical model is described, followed in Section~\ref{sec::num} by the numerical implementation. We discuss the mixed-hybrid-finite-element (MHFE) approximation to fluxes and a pressure equation, and construct CF fracture elements by appropriately integrating over fracture and matrix fluxes. The discontinuous Galerkin (DG) mass transport update, and thermodynamic equilibrium computations are briefly discussed. The numerical implementation includes several improvements over earlier work. We extend the CF equilibrium discrete fracture model to fully compositional three-phase flow and adopt a more stable CFL condition.
In Section~\ref{sec::numex} we provide six numerical examples. Two examples compare the CF model to single-porosity simulations, and the other four examples illustrate features of the model for both the flow of three hydrocarbon phases, and gas-oil-water systems in fractured two- (2D) and three-dimensional (3D) domains.

\section{Mathematical model}
\label{sec::math}
\subsection{Mass transport}
We adopt a fractional flow formulation and write the mass- (or species-) transport equation for the total molar density of each species $i$ in the three-phase mixture as
\begin{equation}\label{eq::transfer}
\phi \frac{\partial c z_i}{\partial t} + \nabla\cdot\vec{U}_i = F_i, \quad i = 1, \ldots, n_c,
\end{equation}
with $\phi$ the porosity and $c$ the overall molar density. For each of $n_c$ components $i$, $z_i$ is the overall molar composition, $F_i$ are sink/source terms representing injection and production wells, and $\vec{U}_i$ is the total molar flux, which consists of \textit{convective} phase fluxes $\vec{\upvartheta}_\alpha$ and \textit{diffusive} phase fluxes $\vec{J}_{i, \alpha}$:
\begin{equation}\label{eq::totmolflux}
\vec{U}_i = \sum_{\alpha} \left(c_\alpha x_{i,\alpha} \vec{\upvartheta}_\alpha + \phi S_\alpha \vec{J}_{i,\alpha}\right),\quad  i = 1, \ldots, n_c.
\end{equation}
The three phases are labeled by $\alpha$, and for each phase,  $ \vec{\vartheta}_\alpha$ is the Darcy flux,  $c_\alpha$ the molar density, and $x_{i,\alpha}$ the mole fraction of component $i$. The phases $\alpha$ can either be oil, gas and water, or three hydrocarbon phases. The reduction of the diffusive flux by the porosity is a minor improvement over the expression in \citet{hoteitdiff} to account for the reduced surface available for diffusion in porous media. The phase diffusive fluxes are weighed by the saturations for similar reasons.
An additional reduction may be included to account for tortuosity.

\subsection{Diffusive fluxes}
The diffusive fluxes $\vec{J}_{i, \alpha}$ are given by 
\begin{subequations}
\begin{eqnarray}\label{eq::diff}
\vec{J}_{i, \alpha} &=& -c_\alpha \sum_{j=1}^{n_c -1} D_{ij, \alpha} \nabla x_{j,\alpha},\quad  i = 1, \ldots, n_c-1,\\\label{eq::nc}
\vec{J}_{n_{c}, \alpha}&=&-\sum_{i=1}^{n_c-1} \vec{J}_{i, \alpha} .
\end{eqnarray}
\end{subequations}
Multi-component Fickian diffusion is modeled by $(n_c - 1)\times(n_c-1)$ matrices of temperature, pressure and composition dependent phase diffusion coefficients, $D_{ij, \alpha}$, and takes into account non-ideality of the fluids. One can easily demonstrate \citep{hoteit2011}, that mass balance is violated when only a single diffusion coefficient, or a diagonal matrix of diffusion coefficient is used, as is generally done in commercial simulators. The off-diagonal components represent dragging effects and can cause a particular species, in a multi-component mixture, to diffuse from a region of low concentration to higher concentration. These effects have been demonstrated experimentally for a three-component system \citep{arnoldtoor}. We calculate the diffusion coefficients from a unified model for open space diffusion \cite{Leahy-Dios}. Dispersive contributions to $D_{ij, \alpha}$ are neglected because they are higher-order in terms of the convective fluxes, which are small in the matrix, and lower-dimensional in the fractures.% 

\subsection{Convective fluxes}
Each of the volumetric convective phase fluxes is given by the corresponding Darcy relation:
\begin{equation}\label{darcy}
\vec{\vartheta}_\alpha = - \lambda_{\alpha}\mathrm{K} (\nabla p - \rho_\alpha \vec{g}),
\end{equation}
in terms of gravitational acceleration $\vec{g}$, permeability tensor $\mathrm{K}$, mobility $\lambda_\alpha (S_\alpha)$, saturation $S_\alpha$, and mass density $\rho_\alpha$. 
A complication of working with individual Darcy phase-fluxes is that \eqref{darcy} cannot be inverted in favor of the pressure when a phase may be absent or immobile. To circumvent this issue, and to reduce the system of equations that has to be solved directly, we adopt the fractional flow formalism and write a Darcy relation for the \textit{total} flux $\vec{\vartheta}_t$. To simplify the notation, we adopt the following definitions for phase mobilities $\lambda_{\alpha}= k_{\mathrm{r}\alpha}/\mu_{\alpha}$, in terms of relative permeabilities $k_{\mathrm{r}\alpha}$ and viscosities $\mu_{\alpha}$, effective phase mobilities $\mathrm{k}_{\alpha} = \lambda_{\alpha} \mathrm{K}$, total (effective) mobilities $\lambda_{t} = \sum_{\alpha} \lambda_{\alpha}$ and $\mathrm{k}_{t} = \sum_{\alpha} \mathrm{k}_{\alpha} $, and fractional flow functions $f_{\alpha} = \lambda_{\alpha}/\lambda_{t}$, and write:
\begin{eqnarray}\label{eq::thetatot}
 \vec{\upvartheta}_{t} &=& \sum_{\alpha}\vec{\upvartheta}_{\alpha}  = -\mathrm{k}_{t} \left( \nabla p - \sum_{\alpha}f_{\alpha} \rho_{\alpha}\vec{g}\right), \end{eqnarray}
The total (effective) mobility is positive definite, so \eqref{eq::thetatot} can be solved for the pressure. As we discuss in detail below, we simultaneously solve for the pressure and for $\vec{\upvartheta}_{t}$. After finding the total flux $\vec{\upvartheta}_{t}$, we can reconstruct the phase fluxes, independent of the pressure, from:
 \begin{subequations}\label{eq::vafromvt}
\begin{eqnarray}
 \vec{\upvartheta}_{\alpha} &=& f_{\alpha} \left(\vec{\upvartheta}_{t} + \vec{G}_\alpha \right),\\
\vec{G}_\alpha &=& \sum_{\alpha^\prime} \mathrm{k}_{\alpha^\prime} (\rho_{\alpha} - \rho_{\alpha^\prime}) \vec{g}.
 \end{eqnarray}
\end{subequations}

\subsection{Pressure equation}
We use Acs's method \citep{acs, watts} to compute the pressure field from: 
\begin{equation}\label{eq::acs}
\phi C_t \frac{\partial p}{\partial t} +  \sum_{i=1}^{n_c} \bar{v}_i (\nabla\cdot \vec{U}_i - F_i) = 0,
\end{equation}
where $C_t$ and $\bar{v}_i$ are, respectively, the total compressibility and total partial molar volumes of the three-phase mixture. Expressions for both variables are derived in Appendix C in \citep{moortgatRSS}. Rock compressibility may also be included in $C_t$.

\subsection{Phase compositions and molar fractions}
Phase compositions $x_{i,\alpha}$ and phase molar fractions $\varpi_\alpha$ are derived from the non-linear set of equations that guarantee equality of fugacities of each component $i$ in all three phases ($\alpha = \alpha_1, \alpha_2, \alpha_3$), as required by local thermodynamic equilibrium. Computationally, the natural logarithm of the equilibrium ratios $K_{i, \alpha}$ is more robust \citep{kj2011}. Selecting one reference phase, say $\alpha_3$, we have two sets of equilibrium ratios $K_{i,\alpha_1} = x_{i,\alpha_1}/x_{i,\alpha_3}$ and $K_{i,\alpha_2} = x_{i,\alpha_2}/x_{i,\alpha_2}$ satisfying the equilibrium criteria:
\begin{subequations}\label{eq::lnk}
\begin{eqnarray}
\ln K_{i,\alpha_1} &=& \ln \varphi_{i,\alpha_3} - \ln \varphi_{i,\alpha_1}, \quad\mbox{and}\\
\ln K_{i,\alpha_2} &=& \ln \varphi_{i,\alpha_3} - \ln \varphi_{i,\alpha_2},
\end{eqnarray}
\end{subequations}
in terms of the fugacity coefficients $\varphi_{i,\alpha}$. \eqref{eq::lnk} is supplemented by the constraint relations:
\begin{subequations}\label{eq::constraints}
\begin{eqnarray}
z_i &=& \sum_\alpha \varpi_\alpha x_{i,\alpha},\quad i = 1, \ldots, n_c - 1, \\
\sum_i z_i &=& \sum_i x_{i,\alpha_1} = \sum_i x_{i,\alpha_2} = \sum_i x_{i,\alpha_3} = 1.
\end{eqnarray}
\end{subequations}

An equation of state has to be specified to describe the three phases and derived quantities, such as saturations, molar and mass densities, compressibilities and partial molar volumes. We model the aqueous phase with a cubic-plus-association EOS that takes into account cross-association between water and $\mathrm{CO}_{2}$ molecules, and self-association of water. In the absence of water, the CPA-EOS reduces to the Peng-Robinson EOS, which we use for pure hydrocarbon phases. We refer the reader to \citep{liCPA, moortgatRSS} for details.

\subsection{Boundary and initial conditions}
To complete the description of the physical model, we prescribe initial and boundary conditions. The initial condition consists of the overall composition and pressure field throughout the domain. 
The boundaries are described by non-overlapping Dirichlet and Neumann conditions: we consider impermeable boundaries, except in production wells where we have either a constant pressure or production rate. Injection wells are placed inside the domain as source terms. Production wells can also be described as sink terms, with impermeable boundaries everywhere.

\section{Numerical model}
\label{sec::num}

\subsection{Mixed hybrid finite element method}\label{sec::mhfe}
\subsubsection{Expansion of convective Darcy flux}
In the MHFE method, the convective and gravitational fluxes are decomposed into their normal components across the edges $E$ of each computational matrix element $K$ as:
\begin{equation}\label{eq::ravthom}
\vec{\upvartheta} (t, \vec{x}) = \sum_{E\in \partial K} q_{K, E}(t) \vec{w}_{K, E} (\vec{x}), \quad\mbox{and}\quad \vec{g}(\vec{x}) = \sum_{E\in \partial K} q^{Kg}_{K, E} \vec{w}_{K, E}(\vec{x}),
\end{equation}  
where $\vec{x} = (x, y)$, $\partial K$ is the boundary of element $K$, and $ \vec{w}_{K, E}$ are the lowest-order Raviart-Thomas basis vector fields \citep{raviartthomas}. These vector functions satisfy the properties
\begin{equation}\label{eq::ww}
\vec{w}_{K,E} \cdot \vec{n}_{K,E^\prime} = \frac{1}{A_{E}}, \quad\mbox{and}\quad \nabla \cdot \vec{w}_{K,E}  = \frac{1}{V_{K}},
\end{equation}
where $A_{E}$ is the length/area of edge/face $E$, $V_{K}$ is the area/volume of element $K$. 
The MHFE weak form of \eqref{eq::thetatot} is obtained by multiplying by $\vec{w}_{K,E}$ and integrating over each element $K$. The pressure gradient term is partially integrated and Gauss' theorem is used, such that we have one volume integral over the pressure, and one surface integral. We define $p_K = \int_K p$ and $\int_{\partial K} p = \sum_E \int_E p = \sum_E tp_{K,E}$, which are the averaged pressure in a matrix element, and the averaged pressures along the element edges/faces. We refer to the latter as pressure traces. The MHFE approximation to Darcy's law can then be written as:
\begin{equation}\label{eq::qmatrix}
q_{K, E} = \theta_{K, E} p_K - \sum_{E^\prime \in\partial K} \beta_{K, E, E^\prime} tp_{K, E^\prime} - \gamma_{K, E}, \quad E\in \partial K.
\end{equation}
The coefficients $\theta_{K, E}$, $\beta_{K, E, E^\prime}$ and $\gamma_{K, E}$ are defined in Appendix~\ref{app::mhfe}.

In two dimensions, fracture elements are initially treated as 1D computational elements, such that volume integrals reduce to line integrals over fracture elements $f$, multiplied by the fracture aperture $\epsilon$. In three dimensions, the fractures are 2D planes with an $\epsilon$ width in the third dimension. Similar to the definitions above, we denote the average pressure in a fracture element by $p_f$, the pressures at the end-points or end-edges $e$ of $\partial f$ by $tp_f$, and the size of fracture element $f$ by $|f|$. The MHFE expression for fracture convective fluxes is:
\begin{equation}\label{eq::qfrac}
q_{f, e} = \theta_{f, e} p_f - \sum_{e^\prime \in\partial f} \beta_{f, e, e^\prime} tp_{f, e^\prime} - \gamma_{f, e}, \quad e\in \partial f.
\end{equation}
However, as discussed in the next section, there is no need to evaluate \eqref{eq::qfrac} explicitly.

\subsubsection{Cross-flow equilibrium approximation}\label{sec::cfe}
\begin{figure}
\centerline{\includegraphics[width=\textwidth]{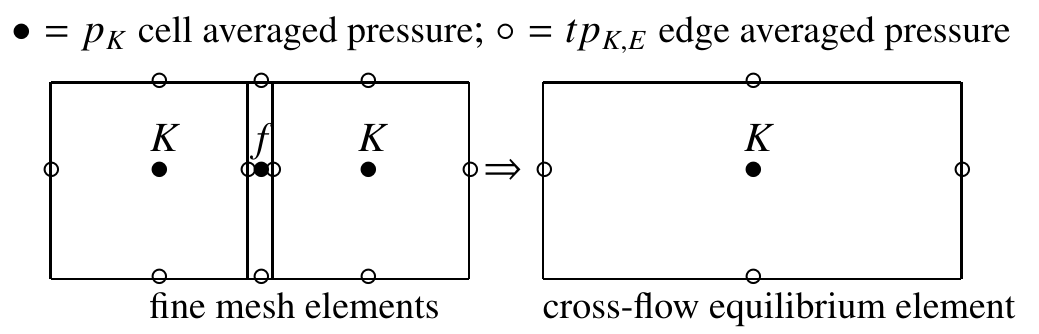}}
\caption{\label{fig::CF}Illustration of cross-flow equilibrium approximation.}
\end{figure}

An explicit treatment of fractures, referred to as single-porosity models, is not computationally feasible. The CFL constraint on the time-step scales with $\Delta t \propto \min(V_{K}/q_K,  V_{f}/q_f)$. For fracture elements, the fracture flux $q_f$ may be high, while the volume $V_{f}$ of a fracture element is generally exceedingly small, particularly for fracture intersections.
To overcome this limitation, we note that the pressure field is continuous, so the pressure in a fracture element is close to the pressure a small distance away in the matrix. We now represent a fracture element, together with two small matrix elements on both sides, as one computational element. This is illustrated in Figure~\ref{fig::CF} for a rectangular 2D mesh. For this larger, combined element, we assign one averaged pressure $p_K$ and $4$ (in 2D) or $6$ (in 3D) pressure traces $tp_{K,E}$. 

Fluxes through edges that are intersected by a fracture (top and bottom edges in Figure~\ref{fig::CF}) are computed by properly integrating over both the matrix and fracture fluxes inside the combined element. We now use $K$ for an element that may contain a fracture, $k$ for the matrix portion of element $K$ and $E$ for an edge that may be intersected by a fracture and write for the total fracture plus matrix flux $\tilde{q}_{K,E}=q_{k, E} + q_{f, e}$, with
\begin{eqnarray}\label{eq:qtot}
\tilde{q}_{K,E} &=&(\theta_{k, E} + \theta_{f, e}) p_K - \sum_{E^\prime \in\partial K} \left(\beta_{k, E, E^\prime} + \beta_{f, e, e^\prime}\delta_{e, E} \delta_{e^\prime,E^\prime}\right) tp_{k, E^\prime} - \gamma_{k, E} - \gamma_{f, e} \delta_{e,E},
\end{eqnarray}
where $\delta_{e,E}$ is $1$ when an edge $E$ is intersected by a fracture element end $e$ and zero otherwise. \eqref{eq:qtot} is similar to the two-phase expressions in \citet{hoteit2006b}. However, in Appendix~\ref{app::mhfe}, we demonstrate that \eqref{eq:qtot} can be more elegantly reduced to \eqref{eq::qmatrix}, with the coefficients $\theta_{K, E}$, $\beta_{K, E, E^\prime}$ and $\gamma_{K, E}$ evaluated in terms of a weighted total effective mobility $\mathrm{k}^\mathrm{eff}_{t}$ across edge $E$:
\begin{equation}\label{eq::keff}
\mathrm{k}^{\mathrm{eff}}_{t} = \mathrm{k}_{t,f} \varepsilon/A_{E}+ \mathrm{k}_{t,m} (1- \varepsilon/A_{E}),
\end{equation}
where $\varepsilon$ is the area of the fracture intersection with $E$, and the subscripts $m$ and $f$ denote matrix and fracture properties, respectively (i.e.~$\mathrm{k}_{t,m}$ is the total effective mobility in the matrix, as defined above \eqref{eq::thetatot}). We emphasize that we allow different relative permeabilities in the fracture and matrix portions of CF elements.

Fluxes in the transverse direction (left and right edges in the figure) are matrix fluxes, computed from \eqref{eq::qmatrix}. The fracture-matrix flux inside the element is accounted for by the assumption that at the fracture-matrix interface there is a large permeability, but a small pressure gradient, which results in a Darcy fracture-matrix flux. The assumption is that this flux instantaneously equilibrates, or mixes, the fluid in the fracture with the fluid in the small neighboring matrix elements (small with respect to the full matrix block). When the MHFE method is combined with a FD mass transport update, this means that the average $c z_i$ for the combined element is updated. For the combination of MHFE with a higher-order DG method, $c z_i$ at the nodes or edges is updated. The mass transport update, \eqref{eq::transfer}, is unaltered from its implementation for homogeneous media. We provide the phase compositions for the combined fracture-matrix element and the summed fluxes, and update the overall molar species densities $c z_i$. We refer to this approach as the cross-flow (CF) equilibrium model and will refer to the combined fracture-matrix elements as CF elements.

In a single-porosity simulation, the CFL condition could be determined by fracture-intersection elements with an area of, say $1\ \mathrm{mm}^{2}$, and fracture fluxes that scale with the high fracture permeability. In the CF approach, when we have, $10\ \mathrm{m}\times 10\ \mathrm{m}$ matrix blocks, we can use CF elements with a width of several $\mathrm{cm}$ or more. This results in a CFL constraint on the time-step that is several orders of magnitude larger than for a single-porosity model. 
A second reason that single-porosity models are computationally expensive, is that the system of equations in the pressure update is ill-conditioned due to the high permeability contrast between the fracture and matrix elements. When we use the averaged CF elements, the pressure update (discussed below) is considerably more efficient.

We emphasize, that the matrix blocks may be discretized by any number of grid-cells, such that we can resolve potential gravitational or viscous fingers in the matrix. The discretization of the matrix blocks is particularly important to model diffusion. The diffusive fluxes are weighed by the phase saturations (\eqref{eq::totmolflux}), and diffusion only occurs within a given phase. When gas is injected in fractured porous media, all the hydrocarbons in the fractures may evaporate into the gas phase, before a gas phase has developed in the neighboring matrix blocks. Because Fickian diffusion only occurs within a phase, one cannot self-consistently compute a diffusive flux from the fractures to the matrix blocks. In reality, the gas and oil at the fracture-matrix interface are in local thermodynamic equilibrium, and dissolution of gas and evaporation of oil occur through Fickian diffusion at the interface. This numerical issue is particularly problematic in single- and dual-porosity models. In the CF approach the problem is alleviated, because the gas in the fractures is mixed with matrix oil and the CF elements may remain in two-phase, particularly in large-scale simulations where breakthrough is avoided. 
When a sufficient amount of light species has accumulated in the neighboring matrix elements, a gas phase may form and diffusion can occur from fracture to matrix in both phases at a high rate. In this fashion, the light species may diffuse element by element into the matrix blocks, while heavier species diffuse towards the fractures. 

Gravitational effects, such as fingering and re-infiltration, where oil drains from a matrix block into a fracture, and then drains from the fracture into another neighboring matrix block, are modeled without special treatment. The CF model is not restricted to sugar-cube fracture configurations, but can be applied to any configuration of discrete fractures in structured or unstructured grids. These features are difficult to incorporate in the dual-porosity model.

\subsubsection{Pressure equation}
The discretization of the pressure equation, \eqref{eq::acs}, is also greatly simplified by the definition of the weighted effective mobility in CF elements. For the convective Darcy flux through a fracture-intersected edge/face $E$, one can easily see that the total phase flux through the matrix and the fracture portions of the CF element reduces to
\begin{equation}\label{eq::fmflux}
\tilde{q}_{\alpha,K,E} = - {\mathrm{k}}^\mathrm{eff}_{\alpha} (\nabla p - \rho_\alpha q^g_{K,E}).
\end{equation}
Similarly, we can use \eqref{eq::vafromvt} with $\mathrm{k}_{\alpha}$ replaced with $\mathrm{k}^{\mathrm{eff}}_{\alpha}$ and $f_{\alpha}$ with $f^{\mathrm{eff}}_{\alpha}$.

We expand the diffusive fluxes, similar to the convective fluxes in \eqref{eq::ravthom} as:
\begin{equation}
\phi S_\alpha  \vec{J}_{i,\alpha} (t, \vec{x}) = \sum_{E\in \partial K} q^\mathrm{diff}_{i,\alpha, K, E}(t) \vec{w}_{K, E} (\vec{x}).
\end{equation}
The diffusion term in \eqref{eq::acs} (with \eqref{eq::totmolflux}) then reduces to $\sum_i \sum_E \bar{\nu}_i  q^\mathrm{diff}_{i,\alpha,K, E}$ and can be combined with the source/sink term in $F_i$. For brevity, we move these terms to the right-hand-side of the equations, denoted as $\mathrm{r.h.s.}$ and define $\zeta = C_t\left(\phi_f V_{f} + \phi_m V_{k} \right)$. The integral form of \eqref{eq::acs} is:
\begin{equation}\label{eq::mhfep1}
\zeta \frac{\partial p_{K}}{\partial t} +  \sum_{i=1}^{n_c}  \bar{v}_{i} \int_{\partial K} (m_{i, K} \vec{\vartheta}_{t,K} - \vec{s}_{i, K})\cdot \vec{n}_K= \mathrm{r.h.s.},
\end{equation}
where in CF elements $\vec{\vartheta}_{t,K}$ is the total flux integrated over both fracture and matrix portions of element $K$, and $m_{i, K}$ and $\vec{s}_{i, K}$ are defined in Appendix~\ref{app::mhfepress}.
In other words, the discretized pressure equation has the same form for fracture-containing CF elements as for matrix elements, i.e.~is the same for fractured and unfractured domains, when written in terms of the weighted effective mobilities and $\zeta $. This formulations is more straightforward than the earlier implementation for two-phase flow in \citet{hoteit2006b}.

Expanding the fluxes as in \eqref{eq::ravthom} and carrying out the integrations using \eqref{eq::ww}, we find
\begin{equation}\label{eq::mhfep2}
\zeta\frac{\partial p_{K}}{\partial t} + \sum_{i=1}^{n_c} \bar{v}_{i}  \sum_{E\in \partial K} (m_{i, K, E} \tilde{q}_{K, E} - \tilde{s}_{i, K, E} ) = \mathrm{r.h.s.}
\end{equation}
We eliminate the fluxes by \eqref{eq::qmatrix} to obtain the spatial discretization of the pressures:
\begin{equation}\label{eq::pdiscr}
\zeta \frac{\partial p_{K}}{\partial t} + \tilde{\alpha}_K p_K - \sum_{E\in K}\tilde{\beta}_{K, E}  tp_{K, E} -  \tilde{\gamma}_K =  \mathrm{r.h.s.}
\end{equation}
For the temporal discretization we use the backward (implicit) Euler method. The fully discretized pressure equation, with the $\mathrm{r.h.s.}$ re-instated, becomes:
\begin{equation}\label{eq::pfinal}
p_K^{n+1} = \frac{\Delta t}{\tilde{\alpha}_K\Delta t + \zeta} \left\{\frac{\zeta}{\Delta t}p^n_K + \sum_{E\in K}\tilde{\beta}_{K, E}  tp_{K, E}^{n+1} +  \tilde{\gamma}_K + 
 \sum_i \bar{\nu}_i\left( F_i - \sum_E  \sum_\alpha q^\mathrm{diff}_{i,\alpha, K, E} \right) \right\},
\end{equation}
with all the coefficients evaluated at the previous time-step.

\subsubsection{Assembly of global matrix for the pressure update}
Eqs.~(\ref{eq::qmatrix}) and (\ref{eq::pfinal}) are for individual elements and edges. To construct the global system of equations to solve, we assume flux continuity across element edges:
\begin{eqnarray}
\tilde{q}_{K, E} + \tilde{q}_{K^\prime, E}  = 0, \quad\mbox{for}\quad E = K \cap K^\prime,
\end{eqnarray}
(note that fluxes are defined with respect to the normal to edge $E$ in element $K$). Collecting the terms for each edge in the domain, we obtain the matrix system
\begin{equation}\label{eq::sysq}
\mathrm{R}^T \vec{P} - \mathrm{M} \vec{T}_P = \vec{I}.
\end{equation}
For $N_K$ grid elements and $N_E$ edges, $\vec{P}$ is the $N_K$-size vector of element averaged pressures, $\vec{T}_P$ is a $N_E$-size vector of the pressure traces at all the edges. The matrices $\mathrm{R}$ and $\mathrm{M}$, and vector $\vec{I}$ are defined in Appendix~\ref{app::global}.

Similarly, we assume pressure continuity, such that at each edge $E$
\begin{equation}
tp_{K,E} = tp_{K^\prime,E}, \quad\mbox{for}\quad E = K \cap K^\prime,
\end{equation}
and construct the matrix system (with matrix definitions given in Appendix~\ref{app::global}):
\begin{equation}\label{eq::sysp}
\mathrm{D} \vec{P} - \tilde{\mathrm{R}} \vec{T}_P = \vec{G}.
\end{equation}
Eqs.~(\ref{eq::sysq}) and (\ref{eq::sysp}) can be combined in one large system that solves simultaneously for the pressures and pressure traces. A more efficient approach presents itself by the fact that $\mathrm{D}$ is diagonal. By multiplying \eqref{eq::sysp} by $\mathrm{D}^{-1}$, we can eliminate $\vec{P}$ from \eqref{eq::sysq} and obtain a system for $\vec{T}_P$ alone:
\begin{equation}\label{eq::finalsys}
(\mathrm{M} - \mathrm{R}^T \mathrm{D}^{-1} \tilde{\mathrm{R}}) \vec{T}_P = \mathrm{R}^T \mathrm{D}^{-1} \vec{G} - \vec{I}.
\end{equation}
\eqref{eq::finalsys} is the system that is solved in the pressure update. The matrix that needs to be inverted has dimensions $N_E\times N_E$, but is sparse. On a structured 2D (3D) grid, each (non-boundary) row/edge has $7$ ($11$) non-zero elements, because $tp_{K,E}$ depends on edge $E$ and the other 3 (5) edges in each of the two neighboring elements. After updating $\vec{T}_p$, $\vec{P}$ is found through inexpensive back-substitution in \eqref{eq::sysp}. The total flux is found by substituting the updated $p_K$ and $tp_K$ in \eqref{eq:qtot}.

The system \eqref{eq::finalsys} is larger than for a FD method, but the considerable advantage is that we simultaneously solve for the element pressures (as in FD), the pressures on the edges (which is advantageous for heterogeneous and fractured domains), and for a continuous velocity field throughout the domain; all with the same order of convergence. 

\subsection{Discontinuous Galerkin mass transport update}
All higher-order methods approximate the mass transport update, \eqref{eq::transfer}, by multiple degrees of freedom for the overall and phase compositions and molar densities. The DG method has the additional advantage that the variables can be  \textit{discontinuous} across edges. One of the benefits is that different orders of approximation can be used in different elements. In our work, we use a bilinear (trilinear) approximation on all 2D (3D) structured elements and a linear approximation on unstructured triangular grids \citep{moortgatI}. Away from strong compositional gradients, we can use a lower-order approximation. Our main interest in the DG method is the simulation of heterogeneous and fractured porous media. At the boundaries between regions of different permeabilities (such as fractures and layers), the phase properties may exhibit strong discontinuities. \textit{Continuous} higher-order methods, such as FD and finite volume, may be used in homogeneous domains, but are a less natural choice to approximate the inherently discontinuous phase properties in fractured reservoirs. 

The higher-order mass transport update reduces numerical dispersion and converges to the exact solution at a higher rate. Alternatively, this means that a given FD result can be reproduced on a significantly coarser grid, and at a correspondingly higher CPU efficiency. Combined with the accurate velocity field from the MHFE method, this approach has been shown to result in orders of magnitude improvement in CPU times \citep{hoteit2005, hoteit2006a, moortgatII, moortgatI, moortgatRSS}. A convergence analysis in \citet{moortgatPetrobras} demonstrates twice the convergence rate for the DG mass transport update as compared to a FD approach.

As was mentioned above, the implementation of the DG method is identical to that in homogeneous domains, which was presented for three-phase flow in \citet{moortgatII} for 2D and \citet{moortgat3D} in 3D, and will not be repeated here. Phase properties are updated at either the edge-centers or the nodes from the total (fracture plus matrix) fluxes through each of the edges. The accuracy is further improved, because we have the pressures at the edges from the MHFE update. To avoid spurious oscillations that may occur in higher-order methods, we use the same slope limiter as in the papers cited above.

\subsection{Phase behavior}
We have developed a phase splitting package that can model both three hydrocarbon phases, with transfer of all species between the three phases, and systems in which one of the phases is water. In the latter case, we neglect the mutual solubility between water and hydrocarbons and water evaporation, and only allow $\mathrm{CO}_{2}$ solubility in water. These are reasonable assumptions for problems where the temperature is below $350\ \mathrm{K}$, and result in considerable computational advantage. 
In the update of phase compositions, first a stability analysis is performed, corresponding to the minimum Gibbs free energy. Then, two- or three-phase-split computations are carried out. When initial guesses are not available, the phase-split routine first performs a number of successive-substitution-iterations (SSI) to obtain a good enough initial guess to switch to the fast-converging Newton method. Generally, the Newton method, based on the natural logarithm of the equilibrium ratios (\eqref{eq::lnk}) only needs one or two iterations to converge. 
When initial guesses are available from the previous time-step, Newton's method is attempted first, without a stability analysis. Various optimizations have resulted in a highly efficient algorithm. The details are provided in an earlier paper on three-phase flow in homogeneous media \citep{moortgatRSS}.

\subsection{Relative permeability and viscosity}\label{sec::relperm}
We adopt Stone I relative permeabilities \citep{stone1, stone2}. In the numerical examples, we will denote the residual saturations for water-oil-gas mixtures by $S_{\mathrm{rw}}^{0}$ for water, $S_{\mathrm{row}}^{0}$ for oil-to-water, $S_{\mathrm{rog}}^{0}$ for oil-to-gas, and $S_{\mathrm{rg}}^{0}$ for gas. Similarly, the end-point relative permeabilities are $k_{\mathrm{rw}}^{0}$, $k_{\mathrm{row}}^{0}$, $k_{\mathrm{rog}}^{0}$, and $k_{\mathrm{rg}}^{0}$, and the powers are $n_{\mathrm{w}}$, $n_{\mathrm{ow}}$, $n_{\mathrm{og}}$, and $n_{\mathrm{g}}$. For mixtures of three hydrocarbon phases, we will use the same notation for simplicity, where the $w$ subscript refers to the third hydrocarbon phase.

We use either the LBC \citep{viscosity} or PC \citep{pedersen} viscosity correlations. Near the critical point, the LBC correlation may perform poorly and the PC correlation, which does not require phase identification, is an improvement. 

\subsection{CFL condition}
Earlier work on higher-order IMPEC modeling of two-phase compositional flow in fractured media \citep{hoteit2005,hoteit2006b,hoteit2009} used the CFL condition for the convective fluxes
\begin{equation}\label{eq::oldcfl}
(\Delta t)_{\mathrm{conv}} \le \min_{K} \left(\frac{\phi |K|}{\sum_{E\in\partial K} \sum_\alpha  |\tilde{q}_{\alpha,K,E}|}\right).
\end{equation}
However, this condition only applies to immiscible flow and does not guarantee that the total number of moles of any species cannot flow out of any grid element in one time-step. Moreover, the summation over the absolute values of the fluxes may result in an overly restrictive time-step when some of the fluxes flow into an element. 

We have implemented a different CFL condition for the compositional fractional flow formulation, equivalent to the suggestion in \citet{coatscfl}, in terms of the convective fluxes:
\begin{equation}\label{eq::newcfl}
(\Delta t)_{\mathrm{conv}} \le \min_{i,K} \left(\frac{\phi |K| c_{K} z_{i,K}}{\sum_{E\in\partial K} \sum_\alpha c_{\alpha,K,E} x_{i,\alpha,K,E} \tilde{q}_{\alpha,K,E}}\right), \quad \forall \tilde{q}_{\alpha,K,E}>0,
\end{equation}
where $c_{\alpha,K,E}$ and $x_{i,\alpha,K,E}$ are, respectively, the phase molar density and composition, evaluated on edge $E$ in the higher-order DG mass transport update. $c_{K}$ and $z_{i,K}$ are the element averaged molar density of the mixture and overall composition in element $K$. By using condition Eq.~\ref{eq::newcfl}, the stability of the algorithm is improved considerably, in particular for fractured media. 

When Fickian diffusion is included, we add the outgoing diffusive fluxes to the denominator in \eqref{eq::newcfl} and check an additional criteria in terms of the maximum eigenvalue of the matrix of diffusion coefficients. Denoting the maximum eigenvalue of each of the matrices of phase diffusion coefficients by $\Lambda_{\mathrm{diff},\alpha,K}$, and $\Lambda_{\mathrm{diff},K} = \max_{\alpha}(\Lambda_{\mathrm{diff},\alpha,K})$, we define
\begin{eqnarray}
(\Delta t)_{\mathrm{diff}} &\le& \min_{K}\left(\frac{\phi |K|}{\Lambda_{\mathrm{diff},K}}\right), \quad\mbox{and}\\
\Delta t &\le& \min((\Delta t)_{\mathrm{conv}}, (\Delta t)_{\mathrm{diff}})
\end{eqnarray}
However, most problems of interest are convection dominated. In homogeneous media, the convective fluxes are generally larger than the diffusive fluxes, even at low injection rates. Fickian diffusion is most pronounced in fractured media, where steep compositional gradients may exist between fractures and matrix blocks. However, the CFL condition is usually determined by the largest convective flux inside small fracture elements. Compared to FD models, the CFL constraint is alleviated, because the higher-order DG method allows the use of coarser grids elements. 

The steps in the full simulator algorithm are similar to other IMPEC codes, and are outlined in more detail in \citet{moortgatI, moortgat3D}.

\section{Numerical examples}\label{sec::numex}
We present six numerical examples to illustrate the strengths of our discrete fracture three-phase flow model. In Examples 1 and 5, we compare cross-flow equilibrium results to single-porosity simulations. In these example, we neglect Fickian diffusion, because single-porosity simulations have a numerical issue in computing diffusion in fractured media, as discussed in Section~\ref{sec::cfe}. 

In Examples 2 and 3, we consider a typical oil recovery scenario for fractured porous media and account for diffusion. The domain is depleted, followed by water flooding and then enhanced oil recovery by $\mathrm{CO}_{2}$ injection. Example 3 considers a high matrix permeability, such that gravitational fingers may develop throughout the domain.
The flow of three hydrocarbon phases in a larger fractured domain is illustrated in 
Example 4. Examples 1 -- 5 are for 2D domains. In the last example, we consider CO$_{2}$ injection into a complex 3D domain with a number of discrete horizontal and vertical fractures. 

\subsection{Example 1: Comparison of discrete CF model to single porosity simulation}\label{sec::ex1}
\begin{figure}
\centerline{
\subfigure[Grid 1]{\includegraphics[height=.6\textheight]{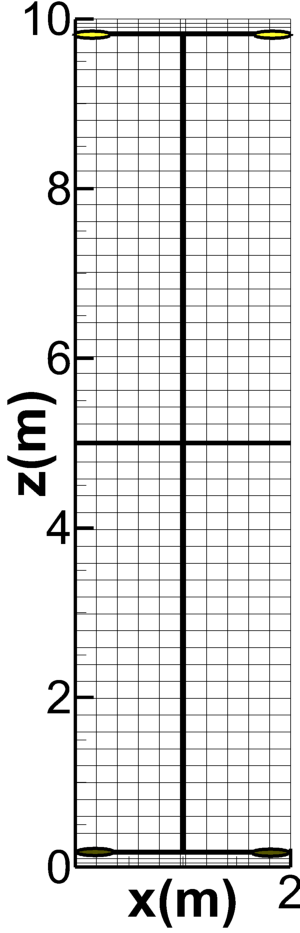}}
\subfigure[Grid 2]{\includegraphics[height=.6\textheight]{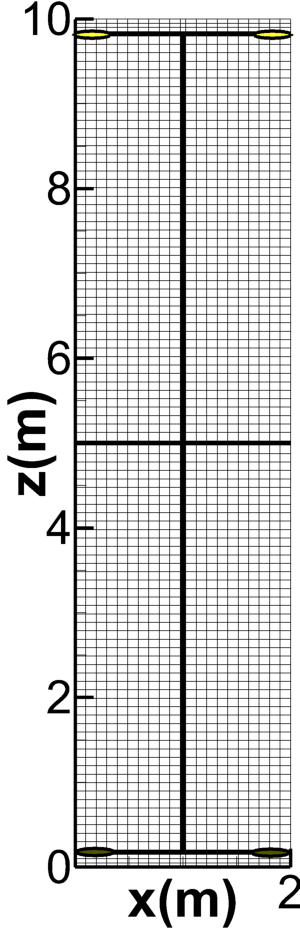}}}
\caption{\label{fig::meshex1} Example 1 --- Computational mesh: $2\ \mathrm{m} \times 10\ \mathrm{m}$ domain, $5\ \mathrm{cm}$ wide CF fracture elements. Grid 1 has $11\times 57$ elements, and grid 2 is $21\times 105$. Injection and production wells are indicated by ellipses near the bottom and top, respectively.}
\end{figure}

We test the CF model by comparing to a single-porosity simulation for water flooding, followed by $\mathrm{CO}_{2}$ injection in a $2\ \mathrm{m} \times 10\ \mathrm{m}$ column with four matrix blocks. Convergence of the CF results is verified by performing simulations on a coarse $11\times 57$ element Grid 1, and a finer $21\times 105$ Grid 2.
The domain, grids and locations of fractures and wells are indicated in Figure~\ref{fig::meshex1}. To make the single-porosity simulation computationally feasible, we assume relatively wide fractures of $5\ \mathrm{mm}$. For the cross-flow simulation we use a $10$ times larger width of $5\ \mathrm{cm}$. The fracture permeability is $4\ \mathrm{d}$, the matrix permeability is $4\ \mathrm{md}$ and matrix porosity is $15\%$.

The column is initial saturated with a light oil, with composition and EOS-parameters given in Table~\ref{table::compex1}. The density and viscosity of the oil, water and $\mathrm{CO}_{2}$ at the initial condition of $T = 350\ \mathrm{K}$, $p_0 = 300\ \mathrm{bar}$ are given in Table~\ref{table::densviscex1}. Both water and $\mathrm{CO}_{2}$ densities are higher than the oil density, and both are injected from the bottom fracture. Production is at constant pressure from the top fracture. The relative permeability parameters for all the examples are listed in Table~\ref{table::relperms}.

\begin{figure}
\centerline{
\subfigure[Single-porosity -- Grid 1]{\includegraphics[height=.6\textheight]{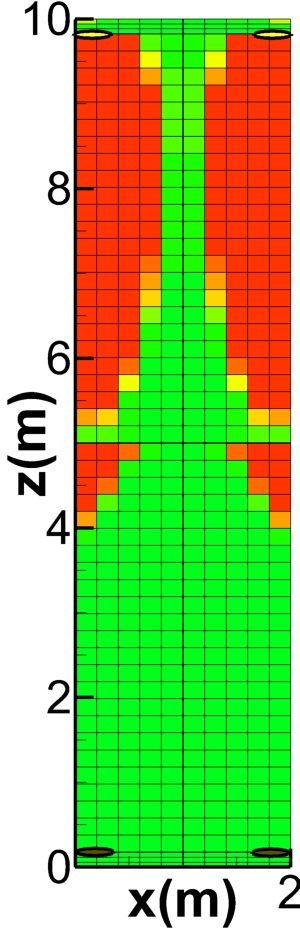}}
\subfigure[CF -- Grid 1]{\includegraphics[height=.6\textheight]{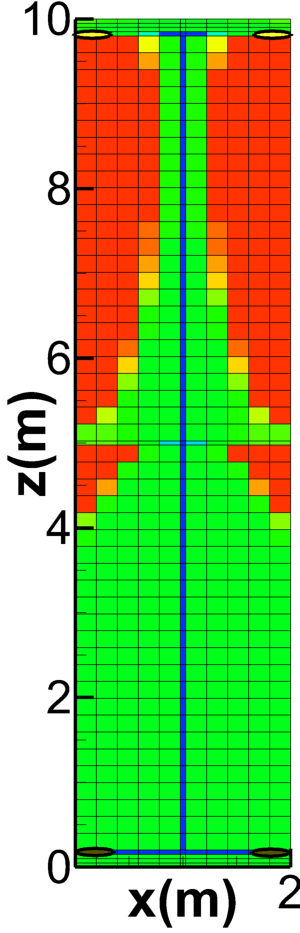}}
\subfigure[CF -- Grid 2]{\includegraphics[height=.6\textheight]{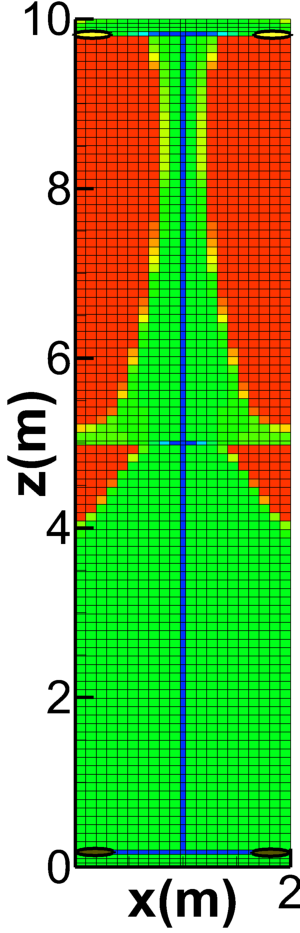}}
\includegraphics[height=.6\textheight]{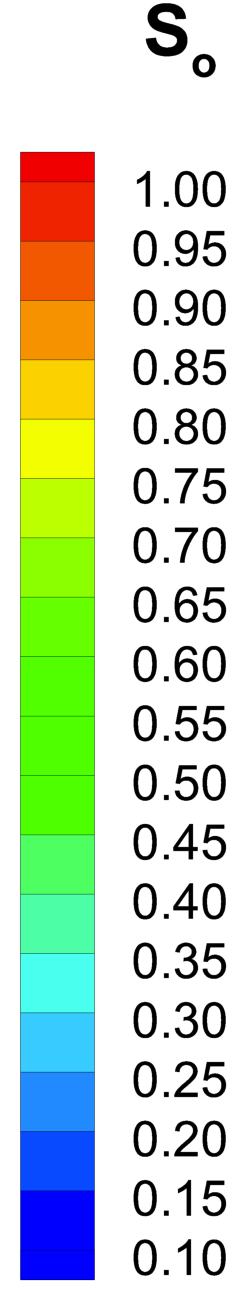}
}
\caption{\label{fig::oilsatsex1100pvi} Example 1 --- Oil saturation at $100\%$ PV water injection for single-porosity simulation on Grid 1, and CF simulations on Grids 1 and 2.}
\end{figure}

\begin{figure}
\centerline{
\subfigure[Single-porosity -- Grid 1]{\includegraphics[height=.6\textheight]{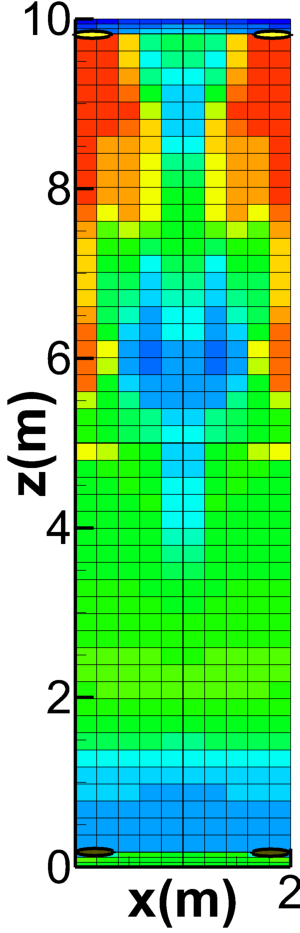}}
\subfigure[CF -- Grid 1]{\includegraphics[height=.6\textheight]{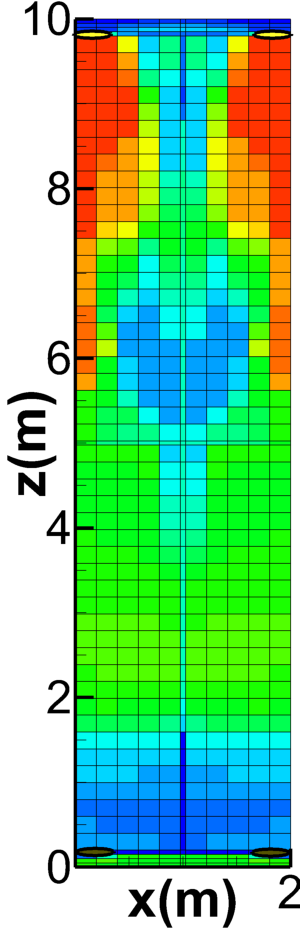}}
\subfigure[CF -- Grid 2]{\includegraphics[height=.6\textheight]{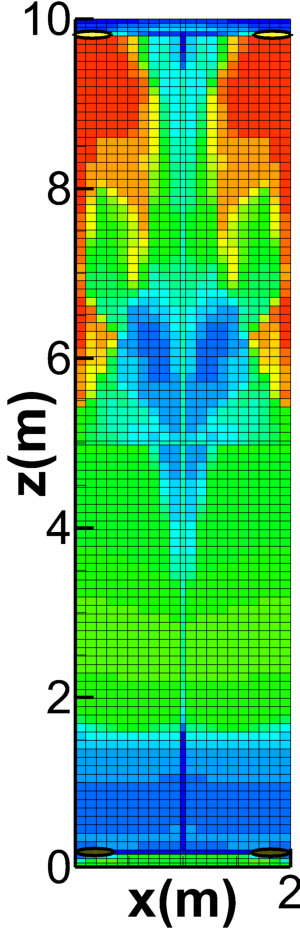}}
\includegraphics[height=.6\textheight]{Fig3d}
}
\caption{\label{fig::oilsatsex1350pvi} Example 1 --- Oil saturation at $100\%$ PV water injection and $250\%$ PV $\mathrm{CO}_{2}$ injection for single-porosity simulation on Grid 1, and CF simulations on Grids 1 and 2.}
\end{figure}

In the simulations, we first inject one pore volume (PV) of water and then $250\%$ PV of $\mathrm{CO}_{2}$ at $2\%$ PV/day. The purpose is to verify that we can obtain the same results with our CF model as with a single-porosity simulation, but at high CPU efficiency. Figure~\ref{fig::oilsatsex1100pvi} shows the oil saturation at the end of water flooding for a single-porosity simulation, and for CF simulations on Grids 1 and 2. The oil saturations at the end of the simulations are given in Figure~\ref{fig::oilsatsex1350pvi}. We find excellent agreement between the single-porosity and CF results, and observe that the CF results have mostly converged on the coarser Grid 1. The CF simulation is $26$ times faster than the single-porosity simulation ($1.23$ hr~\textit{versus} $32$ hrs~on Grid 1). The CFL condition is determined by the grid elements containing fracture intersections, which are about $24$ times larger for the CF discretization ($\mathrm{CFL} \sim 1$ hr) than for the single-porosity elements ($\mathrm{CFL} \sim 2.5$ mins). The lower contrast in permeability between CF and matrix elements also results in a more efficient pressure update.
Note that for a smaller fracture aperture the difference in CPU time would be orders of magnitude, and the single-porosity simulation would not be feasible, even for this small problem. 

\begin{figure*}
\centerline{
\includegraphics[width=.5\textwidth]{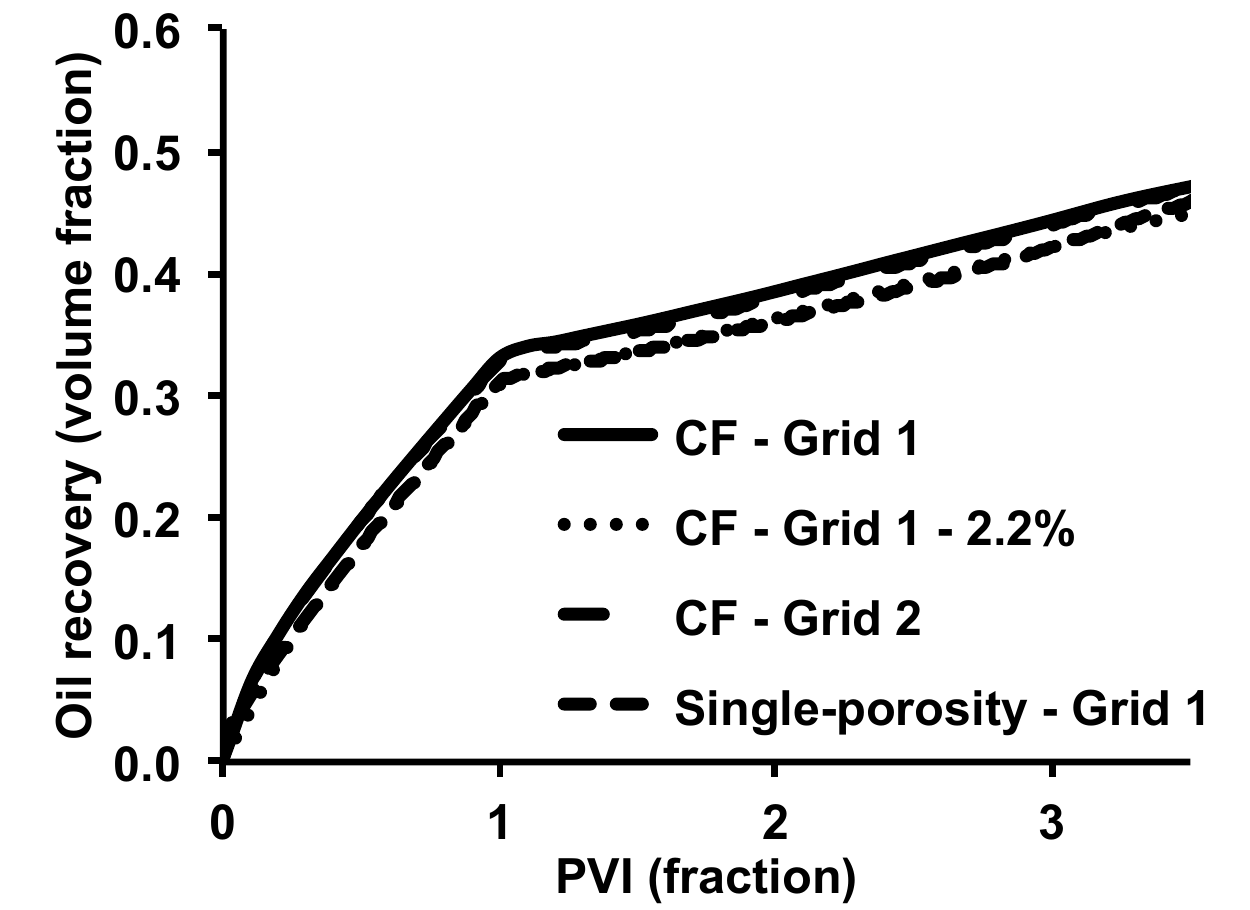}} 
\caption{\label{fig::ex1recovery} Example 1 --- Oil recovery.}
\end{figure*}

Figure~\ref{fig::ex1recovery} shows the oil recovery for both simulations. The results are close, but the recovery from the CF simulation is about $2.2\%$ higher. The reason is that when the gas in the fractures is mixed with a small amount of matrix fluid, the light oil evaporates and is quickly recovered in this small-scale example, where breakthrough occurs early. When we subtract the small pore volume of oil included in the CF elements, the recoveries from CF and single-porosity simulations are in perfect agreement. Also shown is that the CF results on Grids 1 and 2 have converged in terms of oil recovery predictions (the CPU time for the CF simulation on Grid 2 is $3$ hrs).

Further verification of the CF model for single- and two-phase flow was made in \citet{hoteit2005,hoteit2006b}. Now that we have confidence in the CF model, we proceed to a similar problem, but on a large scale that is computationally too expensive for a single-porosity simulation.

\subsection{Example 2: Depletion, water flooding and $\mathrm{CO}_{2}$ injection in fracture porous media}
\begin{figure}
\centerline{
\subfigure[Grid 1 -- $39\times 39$]{\includegraphics[width=.33\textwidth]{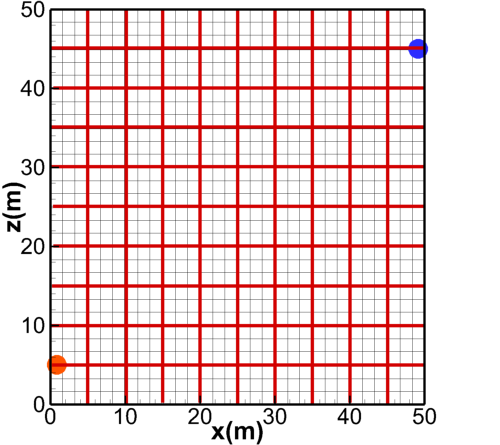}}
\subfigure[Grid 2 -- $49\times 49$]{\includegraphics[width=.33\textwidth]{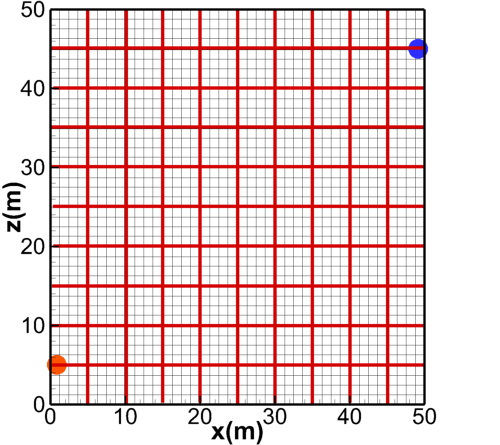}}
\subfigure[Grid 3 -- $79\times 79$]{\includegraphics[width=.33\textwidth]{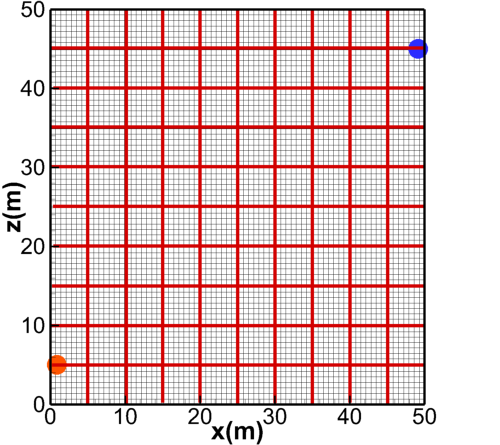}}
}
\caption{\label{fig::mesh} Example 2 --- Computational grids for $50\ \mathrm{m} \times 50\ \mathrm{m}$ domain, $5\ \mathrm{m} \times 5\ \mathrm{m}$ matrix blocks, $20\ \mathrm{cm}$ wide CF fracture elements. Injection and production wells are indicated by circles.}
\end{figure}

We consider a $50\ \mathrm{m}\ \times\ 50\ \mathrm{m}$ domain with $5\ \mathrm{m}\ \times\ 5\ \mathrm{m}$ matrix blocks. Again, we verify convergence on three different mesh refinements. The grids, fractures and wells are shown in Figure~\ref{fig::mesh}, and results will be illustrated on Grid 3, unless stated otherwise. The matrix-blocks have a porosity of $20\%$ and permeability of $10\ \mathrm{md}$. The fractures have an aperture of $1\ \mathrm{mm}$ and permeability of $25\ \mathrm{d}$, and the width of the CF fracture elements is $20\ \mathrm{cm}$. 
The domain is initially saturated with oil, with composition and EOS parameters given in Table~\ref{table::fluidtor}. The temperature is $350\ \mathrm{K}$ and at this temperature, the fluid has a bubble point pressure of $338\ \mathrm{bar}$. 

The simulation is started at an initial pressure of $350\ \mathrm{bar}$ at the bottom of the domain. At this pressure, the entire domain is in single-phase. 
In the first recovery phase, we deplete the domain at a constant rate of $5\%$ PV/yr (computed at the initial pressure) for one year, until the pressure has dropped to $300\ \mathrm{bar}$. Below the bubble point, a gas cap develops in the top of the domain. Table~ \ref{table::densvisc300bar} provides the densities and viscosities for water, oil, $\mathrm{CO}_{2}$, and liberated gas at $T = 350\ \mathrm{K}$ and $p = 300$ and $350\ \mathrm{bar}$. The water and $\mathrm{CO}_{2}$ densities are higher than the oil density, so both are injected from the bottom.

\begin{figure}
\centerline{\includegraphics[width=.5\textwidth]{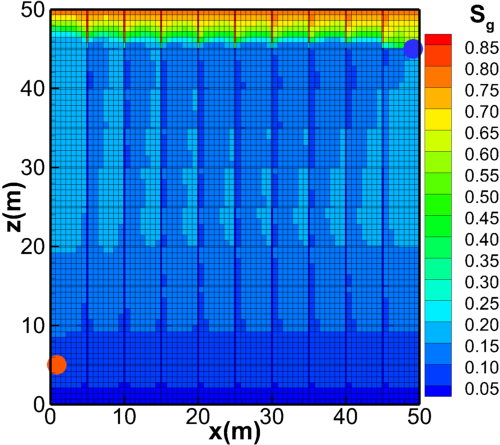}
\includegraphics[width=.5\textwidth]{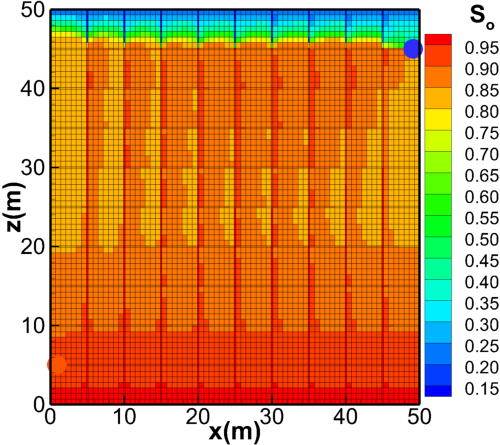}
}
\caption{\label{fig::depletionsats} Example 2 --- Gas (\textbf{left}) and oil (\textbf{right}) saturation after depleting $5\%$ PV in one year on Grid 3.}
\end{figure}

\begin{figure*}
\centerline{
\includegraphics[width=.5\textwidth]{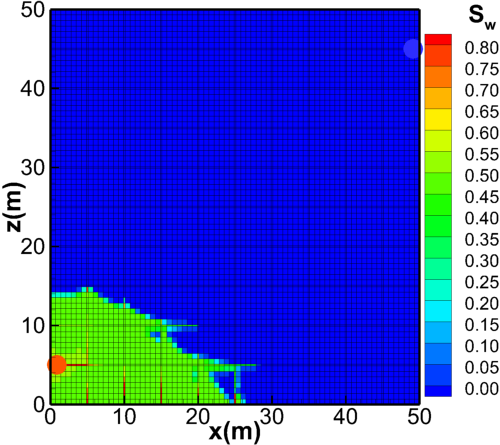}
\includegraphics[width=.5\textwidth]{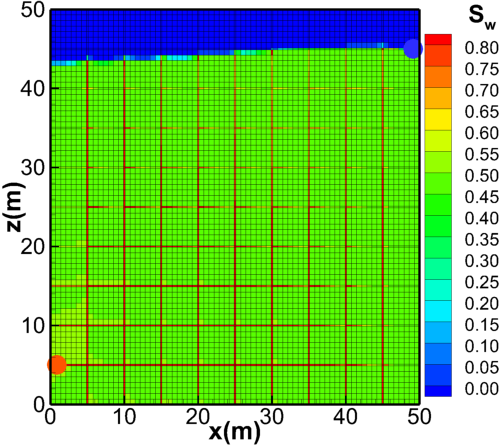}
}
\caption{\label{fig::watflood} Example 2 --- Water saturation at $5\%$ (\textbf{left}) and $50\%$ (\textbf{right}) PV of water-flooding on Grid 3.}
\end{figure*}

\begin{figure*}
\centerline{
\subfigure[$\mathrm{CO}_{2}$]{\includegraphics[width=.5\textwidth]{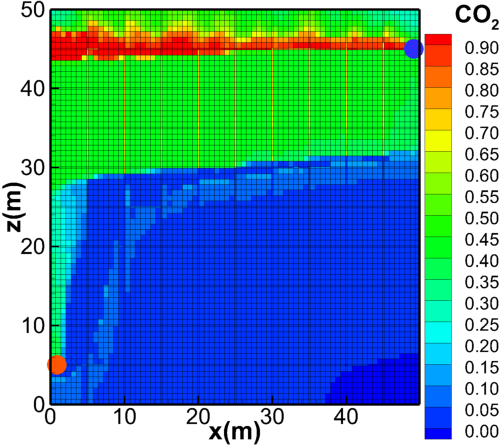}}
\subfigure[$\mathrm{CO}_{2,g}$]{\includegraphics[width=.5\textwidth]{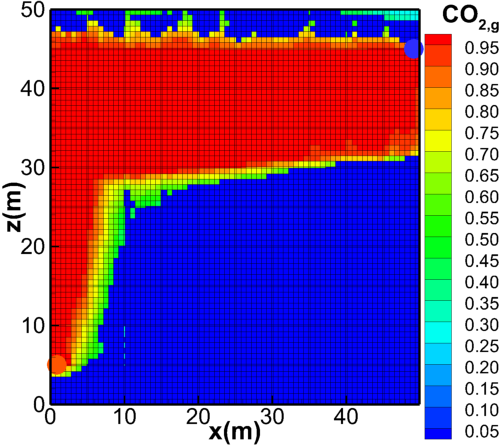}}
}
\centerline{
\subfigure[$\mathrm{CO}_{2,o}$]{\includegraphics[width=.5\textwidth]{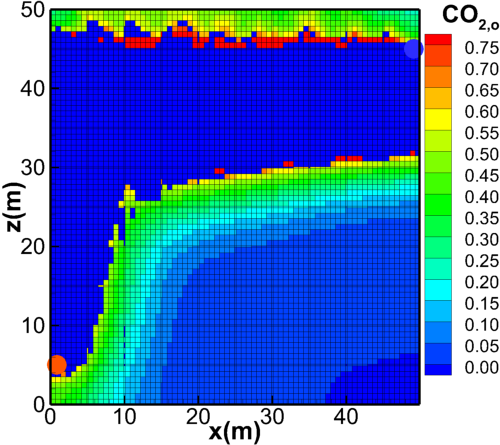}}
\subfigure[$\mathrm{CO}_{2,w}$]{\includegraphics[width=.5\textwidth]{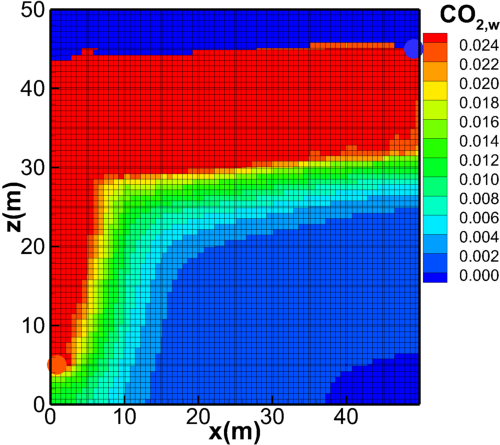}}
}
\caption{\label{fig::CO2comps} Example 2 --- Overall, gas, oil and water molar composition of $\mathrm{CO}_{2}$ at $100\%$ PVI of $\mathrm{CO}_{2}$ on Grid 3.}
\end{figure*}

Figure~\ref{fig::depletionsats} shows the gas and oil saturations at the end of the depletion stage. The gas, liberated at the reduced pressure, has segregated to the top through the high-permeability fractures. 
During secondary recovery, $50\%$ PV of water is injected at a constant rate of $5\%$ PV/yr from the bottom well and production is at a constant pressure from the  top. 
Figure~\ref{fig::watflood} shows the water saturation at $5\%$ and $50\%$ PVI. Because of the residual oil saturation to water, breakthrough has occurred and further water injection is inefficient. 
To recover the residual oil, we consider tertiary recovery by $\mathrm{CO}_{2}$ injection for $20$ years at $5\%$ PV/yr. 
The overall and three phase compositions of $\mathrm{CO}_{2}$ at one PV of $\mathrm{CO}_{2}$-injection are shown in Figure~\ref{fig::CO2comps}. In particular, we note the $\mathrm{CO}_{2}$ dissolution in the aqueous phase in the large three-phase region. $\mathrm{CO}_{2}$ has a solubility of $2.5$ mol\%, or about $6$ wt\% at the given pressure and temperature. The $\mathrm{CO}_{2}$ that dissolves in water is lost to the oil sweep, but results in swelling of the aqueous phase by $\sim 5\%$. At the same time, $\mathrm{CO}_{2}$ has a high solubility in oil which leads to swelling of the oil phase as well. 

\begin{figure*}
\centerline{
\includegraphics[width=.5\textwidth]{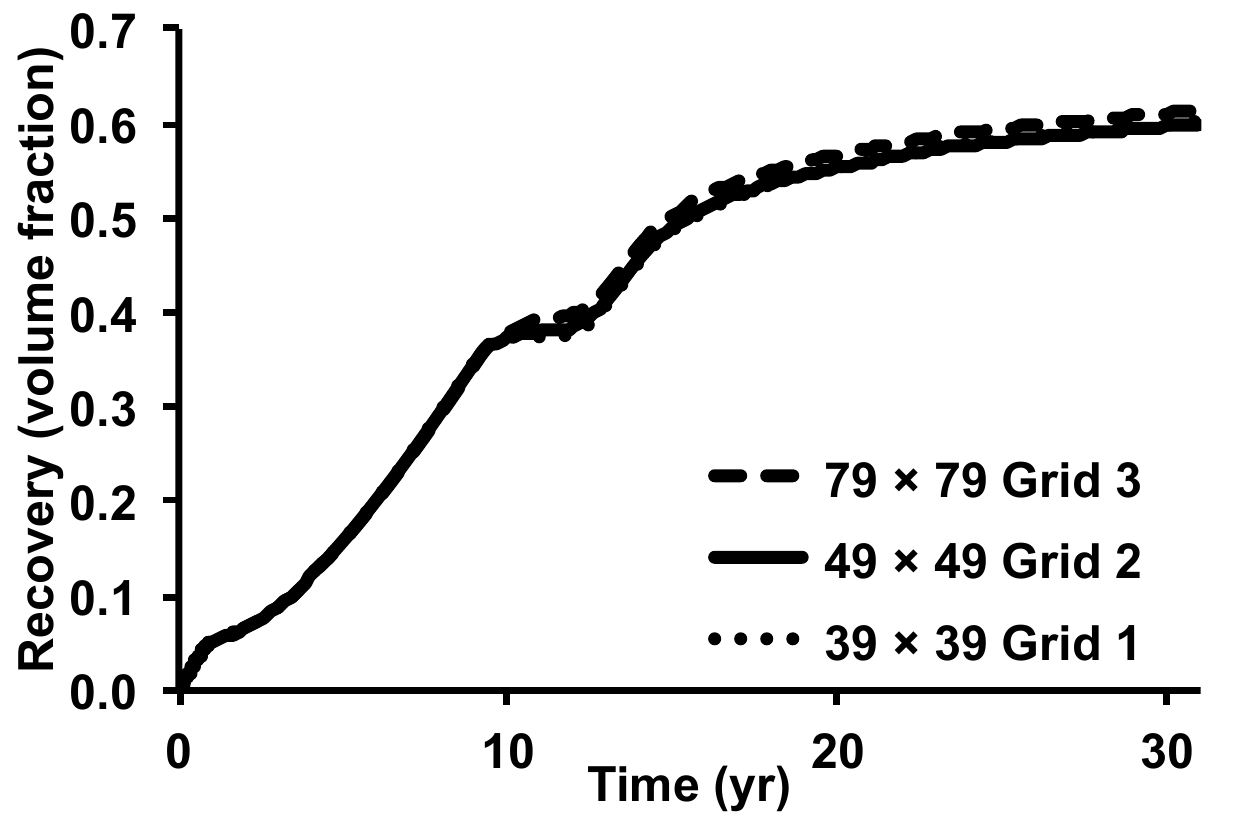}
 } 
\caption{\label{fig::ex2recovery} Example 2 --- Oil recovery.}
\end{figure*}

Figure~\ref{fig::ex2recovery} shows the oil recovery from depletion, water flooding and $\mathrm{CO}_{2}$ injection. Oil recovery from depletion is $5\%$. Water flooding produces another $30\%$ recovery and the enhanced oil recovery by $\mathrm{CO}_{2}$ injection achieves a final recovery to $60\%$. 
Figure~\ref{fig::ex2recovery} also shows that the DG results have converged even on the coarsest mesh, in which the matrix blocks are discretized by only three elements in each direction. 
The CPU times for this simulation are $10$, $19$ and $34$ hrs for Grid 1, 2 and 3, respectively. Because the domain is fractured, phase states and compositions vary wildly throughout the domain and $95\%$ of the CPU time is consumed by the phase split computations. As an illustration of the efficiency of the CF model itself: without the phase-split calculations, the CPU time on the finest mesh is only $2.5$ hrs. The computational cost of three-phase split computations in fractured media motivates continued efforts in improving the efficiency of these algorithms.

\subsection{Example 3: Gravitational fingering in high-permeability porous media}
We repeat the example above but increase the matrix permeability by a factor $10$, and consider a temperature of $400$ K, such that the $\mathrm{CO}_{2}$ is lighter than the oil (Table~ \ref{table::densvisc300bar}). After depletion and water flooding, $\mathrm{CO}_{2}$ is injected from the top well, and production is at constant pressure from the bottom. The simulations are carried out on Grid 2 in Figure~\ref{fig::mesh}, and account for Fickian diffusion.

When $\mathrm{CO}_{2}$ dissolves in water, it increases the density of the aqueous phase by about $1\%$. Because $\mathrm{CO}_{2}$ is injected on top of the previously injected water, the density increase first occurs in the top. Even a density change this small may be gravitationally unstable and trigger a fingering instability. The dissolution of $\mathrm{CO}_{2}$ in oil results in a similar density increase in single-phase, and a higher density increase when light components evaporate from the oil into the $\mathrm{CO}_{2}$-rich gas phase. When the matrix permeability is low, the gravitational fingers do not develop and may be stabilized by Fickian diffusion. At higher permeability, the fingering speed scales linearly with permeability.

\begin{figure*}
\centerline{
\includegraphics[width=.5\textwidth]{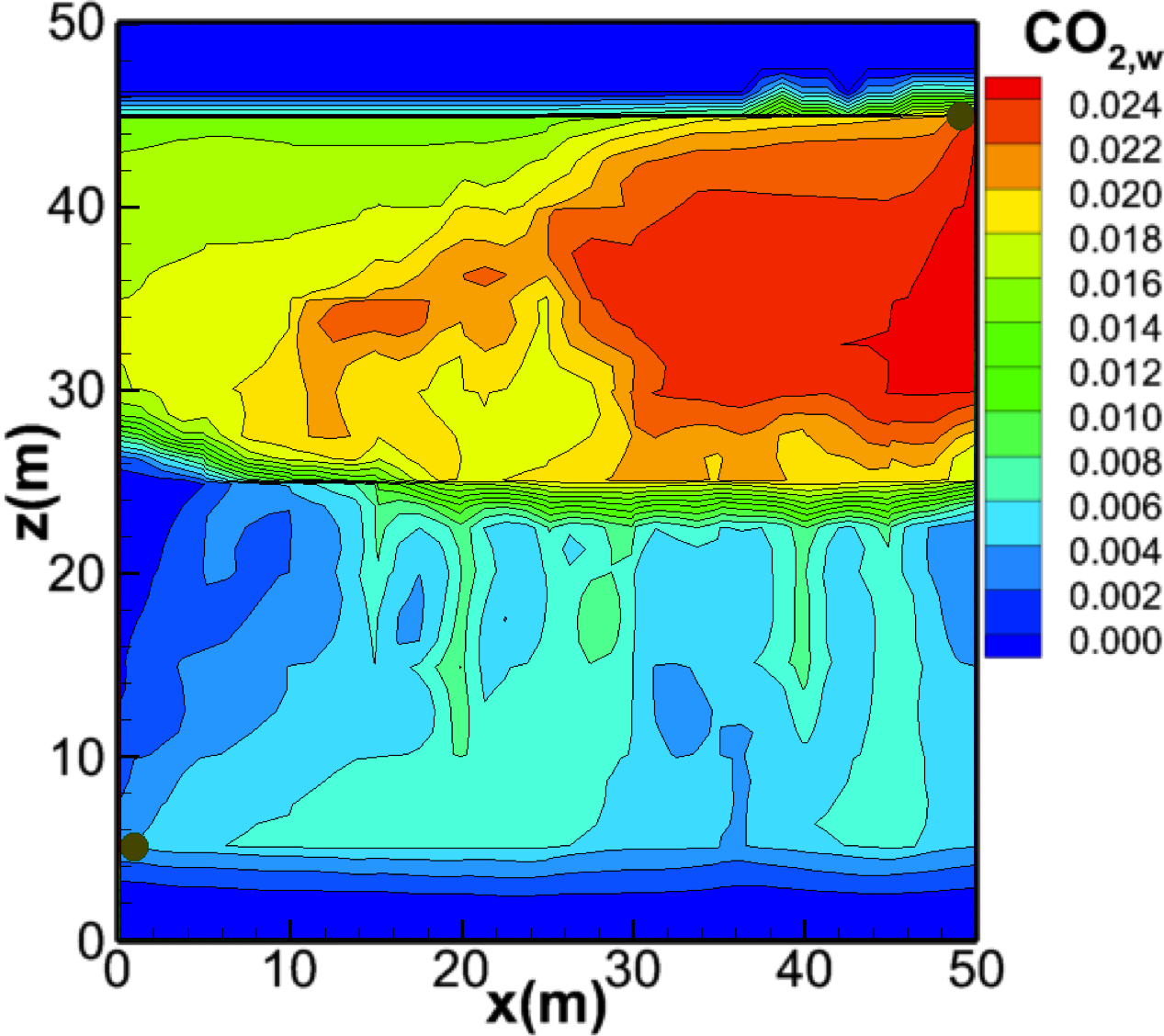}
\includegraphics[width=.5\textwidth]{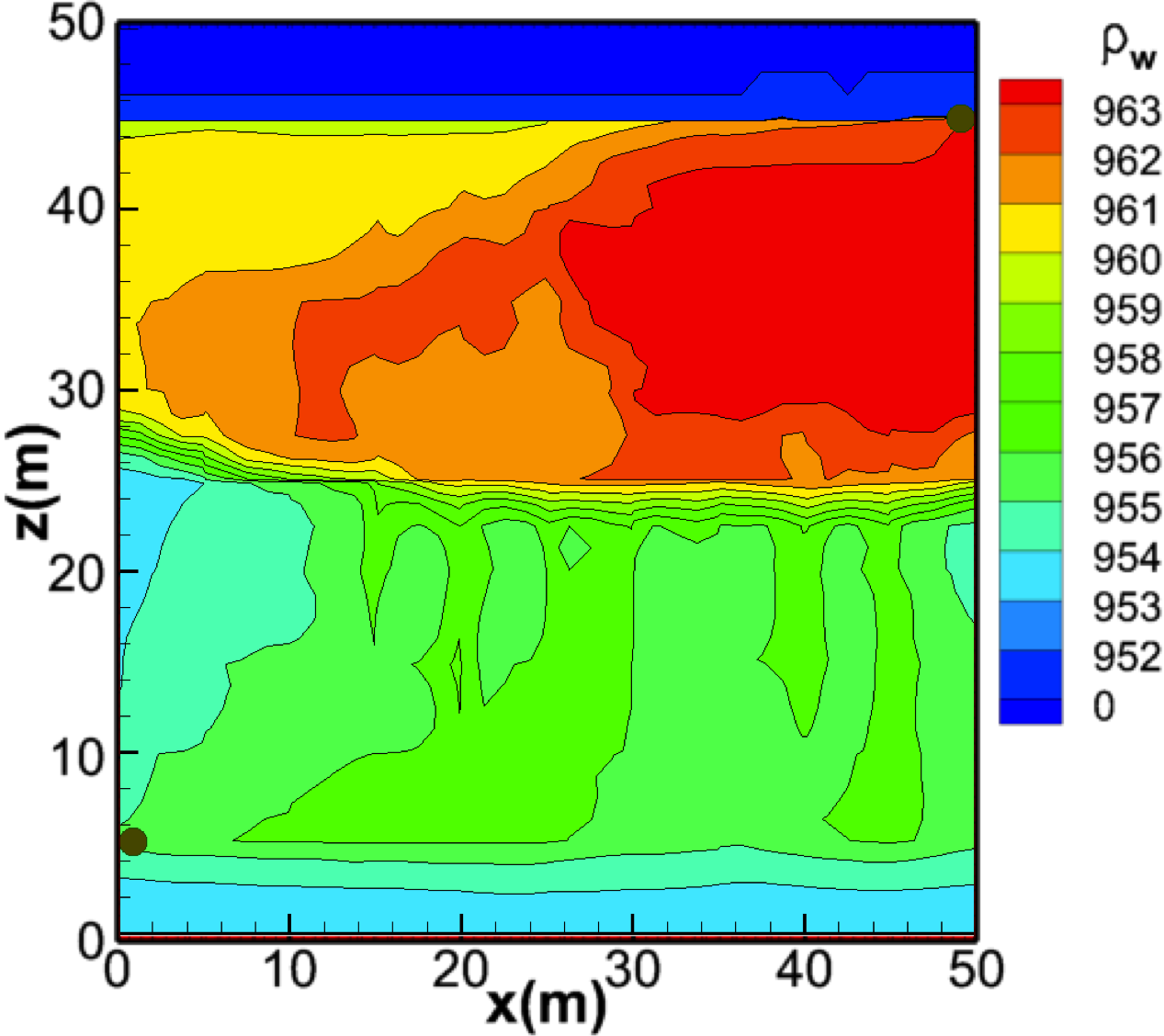}}
\caption{\label{fig::highperm} Example 3 --- $\mathrm{CO}_{2}$ molar composition in aqueous phase (\textbf{left}) and aqueous phase density in kg/m$^{3}$ (\textbf{right}) at $30\%$ PV of $\mathrm{CO}_{2}$ injection, with $\mathrm{K}_m = 100\ \mathrm{md}$ on Grid 2.}
\end{figure*}

Figure~\ref{fig::highperm} shows the $\mathrm{CO}_{2}$ composition in the aqueous phase and resulting density increase after depletion, water flooding and injection of $30\%$ PVI $\mathrm{CO}_{2}$. Pronounced gravitational fingers have developed that span multiple matrix blocks and fractures. Complicated flow patterns like this may not be studied with dual-porosity models. In FD simulations, gravitational and viscous instabilities are often suppressed by numerical dispersion, unless unreasonably fine meshes are used. This example illustrates many of the powerful features or our model: the density increase of water, predicted by the CPA-EOS in three-phase flow, the high accuracy and low numerical dispersion of our higher-order finite element methods, and the efficient modeling of fractures (CPU time of $5.3$ hrs).

\subsection{Example 4: Flow of three hydrocarbon phases in fractured porous media}
We consider the flow of three hydrocarbon phases in a fractured domain, with transfer of all species between the three phases, but neglecting diffusion. The domain, fracture network and aperture, and porosity are the same as Grid 2 in the previous two examples. The matrix and fracture permeabilities are $4\ \mathrm{md}$ and $40\ \mathrm{d}$, respectively. 
The domain is initially saturated with the North Ward Estes oil, and the fluid composition and EOS-parameters can be found in \citet{moortgatRSS, khan, okuno}. The temperature is $301\ \mathrm{K}$, the initial pressure at the bottom is $75\ \mathrm{bar}$, and the relative permeability data are provided in Table~\ref{table::relperms}.

\begin{figure*}
\centerline{
\subfigure[$\mathrm{CO}_{2}$]{\includegraphics[width=.5\textwidth]{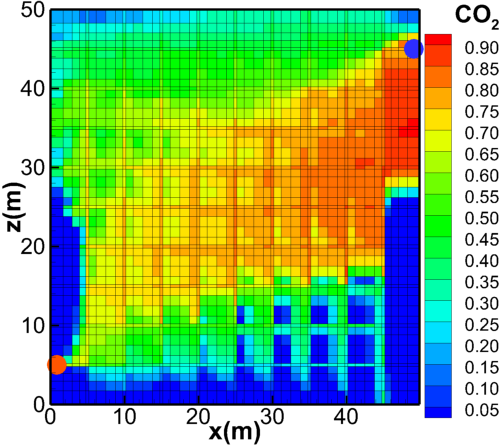}}
\subfigure[$\mathrm{CO}_{2,g}$]{\includegraphics[width=.5\textwidth]{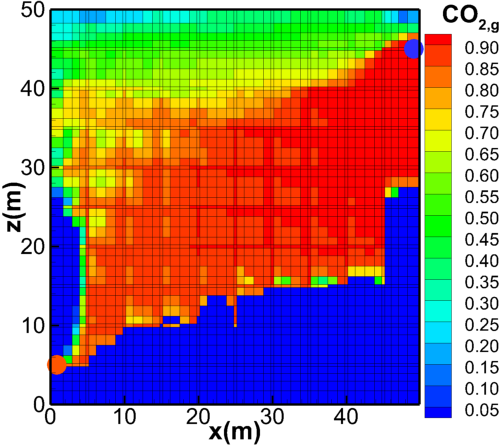}}
}
\centerline{
\subfigure[$\mathrm{CO}_{2,o}$]{\includegraphics[width=.5\textwidth]{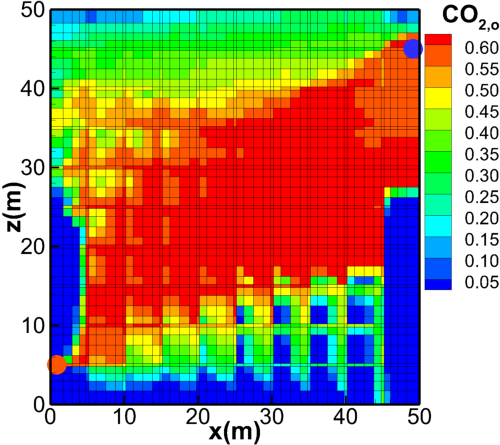}}
\subfigure[$\mathrm{CO}_{2,i}$]{\includegraphics[width=.5\textwidth]{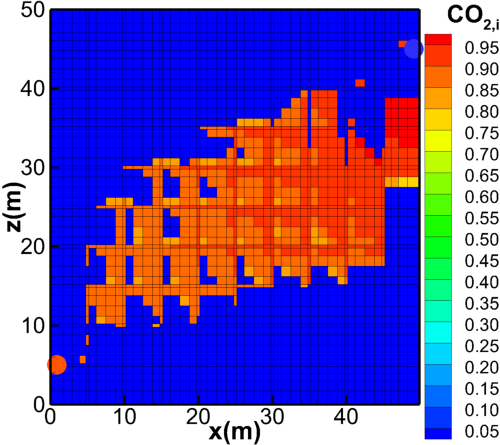}}
}
\caption{\label{fig::CO2comps100pviex3} Example 4 --- Overall, gas, oil and intermediate-phase molar composition of $\mathrm{CO}_{2}$ at $100\%$ PVI on Grid 2.}
\end{figure*}

We inject a mixture of $95$ mol\% $\mathrm{CO}_{2}$ and $5$ mol\% methane from the top at $10\%$ PV/yr for 10 years, and produce at a constant pressure from the bottom.  
Figure~\ref{fig::CO2comps100pviex3} shows the overall and three phase molar compositions of $\mathrm{CO}_{2}$ at $100\%$ PVI. In the region between the lightest $\mathrm{CO}_{2}$-rich gas phase in the top and densest oil phase in the bottom, a large intermediate $\mathrm{CO}_{2}$-rich phase has developed (denoted by $\mathrm{CO}_{2,i}$). This third phase has a low residual saturation ($5\%$) and is readily recovered, while the oil phase is stripped from some of its lighter components and remains as a denser more viscous residual oil. 
Such flow properties can not be captured by a two-phase compositional simulator. Moreover, we find, in line with remarks in \citet{okuno}, that modeling an intrinsically three-phase problem with a two-phase simulator may result in erratic behavior and crashes. In three-phase regions, the phase compositions obtained from a two-phase-split computation do not correspond to the lowest Gibbs free energy (which is the three-phase result). The derived phase properties may be unstable to small perturbations. 

\begin{figure*}
\centerline{
\includegraphics[width=.5\textwidth]{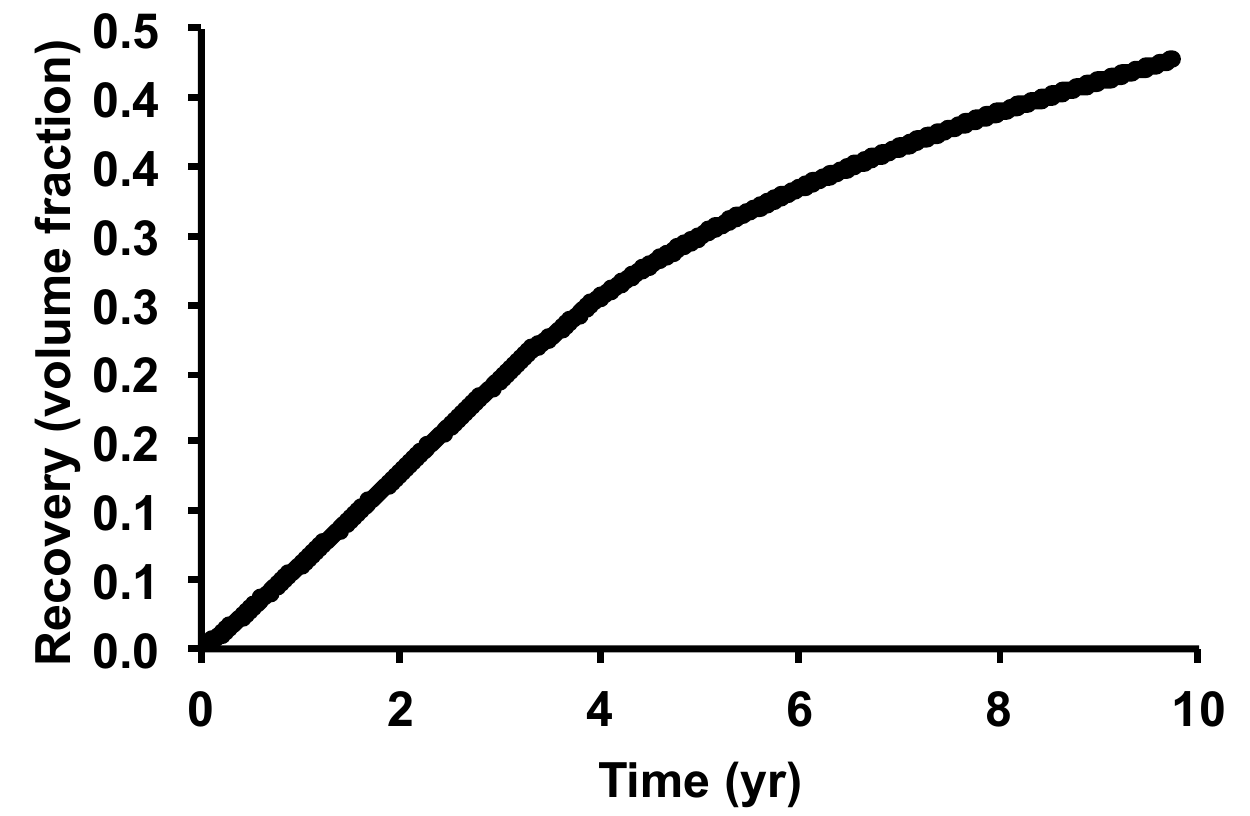}
 } 
\caption{\label{fig::ex3recovery} Example 4 --- Oil recovery.}
\end{figure*}

The oil recovery for this example is provided in Figure~\ref{fig::ex3recovery}. The recovery is not intended to be representative for a field scale project, in which $\mathrm{CO}_{2}$ injection should only be considered for a shorter duration, and certainly not long beyond breakthrough.

The total CPU time for this example is $6\ \mathrm{hr}$, the average CFL time-step is $5\ \mathrm{hr}$, the CPU time per time-step is one second, with $97\%$ of the computation time spent on the phase-split computations.  As was mentioned above, the cost of phase-split computations in fractured media is considerably higher than for homogeneous domains, because phase boundaries occur throughout the domain. Away from phase boundaries, most phase-split computations can be avoided by criteria to determine whether an element was and remains in single-phase. Stability analyses can be skipped in two- and three-phase regions when initial guesses are available from the previous time-step. In fractured media, compositions vary significantly throughout the domain, and full stability and phase-split computations have to be carried out in most elements. For this example, $14\%$ of three-phase split computations were carried out with initial guesses from the previous time-step, avoiding the stability analysis and two-phase split, and only using a couple of Newton iterations. Another $15\%$ of phase-split calculations were avoided by determining which elements were and remained in single-phase. As an indication of the efficiency of the MHFE+DG and CFE fracture model, the CPU time for this example, with the phase-split computations subtracted, is only $8\ \mathrm{mins}$. The large fraction of CPU time spent on the phase-split computations is of course partly due to the small mesh size of only 2401 elements, for which the pressure and mass transport updates are extremely fast. We also note that we compute phase compositions to an accuracy of $10^{-10}$, and the final mass balance error for each of the components is of the order of $10^{-15}$.

\subsection{Example 5: Comparison of CF and fine mesh simulations for three hydrocarbon phases}\label{sec::ex5}
\begin{figure}
\centerline{\includegraphics[width=.7\textwidth]{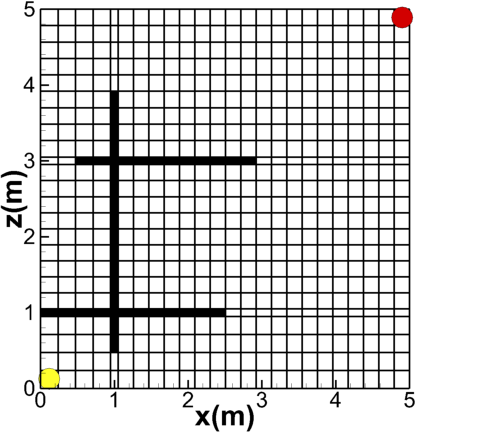}}
\caption{\label{fig::meshex4} Example 5 --- Computational mesh: $5\ \mathrm{m} \times 5\ \mathrm{m}$ domain, $10\ \mathrm{cm}$ wide CF fracture elements, $24\times 24$ grid elements. Injection and production wells are indicated by circles in the bottom-left and top-right corners, respectively.}
\end{figure}

We compare briefly the CF model to a single-porosity simulation for a three hydrocarbon phase fully compositional problem. The  parameters are the same as in the previous example. To compare to a single-porosity simulation, we consider a smaller $5\ \mathrm{m}\ \times\ 5\ \mathrm{m}$ domain with three discrete fractures, as indicated in Figure~\ref{fig::meshex4} for CF elements with a large width of $10\ \mathrm{cm}$. The fracture aperture is $4\ \mathrm{mm}$. Gas, with the same composition as in the previous example, is injected from the bottom at $10\%$ PV/yr and production is at constant pressure from the top.

\begin{figure*}
\centerline{
\subfigure[Fine mesh -- $\mathrm{CO}_{2,g}$]{\includegraphics[width=.4\textwidth]{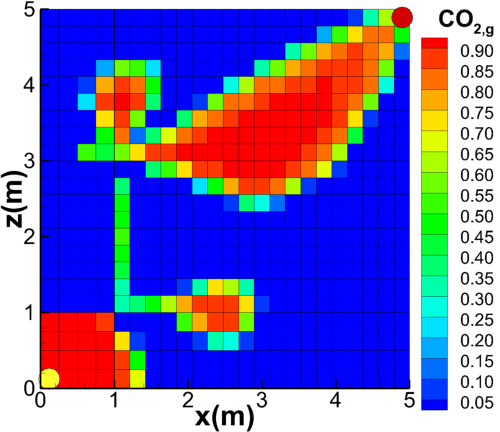}}
\subfigure[CF -- $\mathrm{CO}_{2,g}$]{\includegraphics[width=.4\textwidth]{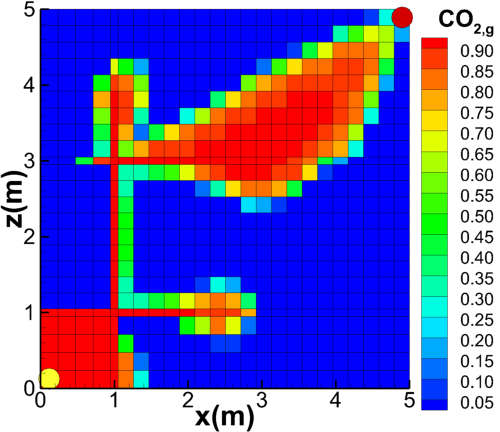}}
}
\centerline{
\subfigure[Fine mesh -- $\mathrm{CO}_{2,o}$]{\includegraphics[width=.4\textwidth]{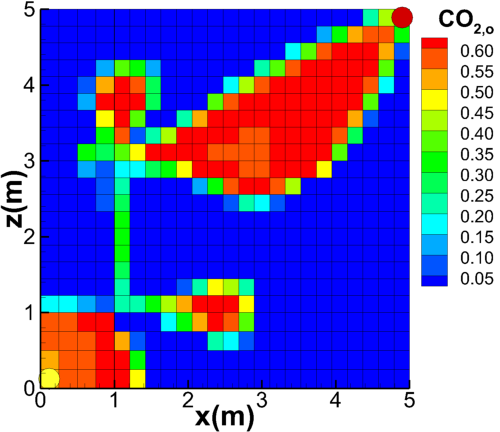}}
\subfigure[CF -- $\mathrm{CO}_{2,o}$]{\includegraphics[width=.4\textwidth]{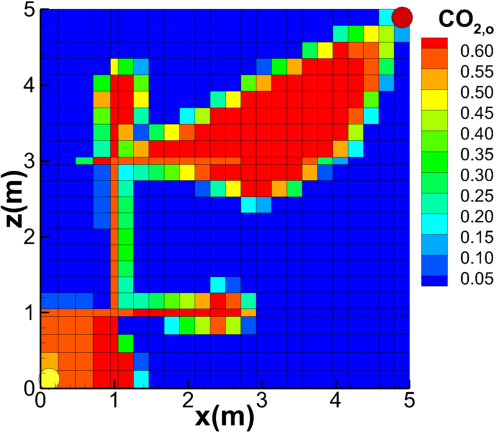}}
}
\centerline{
\subfigure[Fine mesh -- $\mathrm{CO}_{2,i}$]{\includegraphics[width=.4\textwidth]{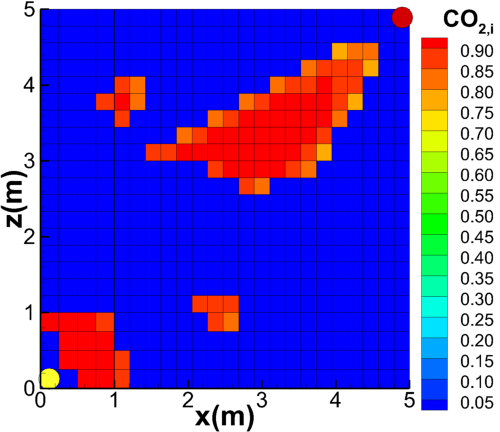}}
\subfigure[CF -- $\mathrm{CO}_{2,i}$]{\includegraphics[width=.4\textwidth]{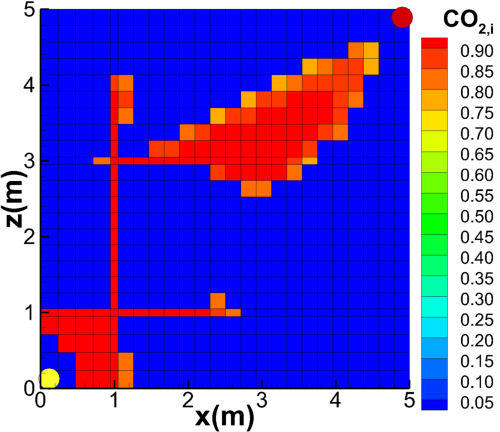}}
}
\caption{\label{fig::CO2compsex4} Example 5 --- Gas, oil and intermediate-phase molar composition of $\mathrm{CO}_{2}$ at $30\%$ PVI.}
\end{figure*}

Figure~\ref{fig::CO2compsex4} shows the three phase compositions of $\mathrm{CO}_{2}$ at $30\%$ PVI. There are small differences, but the main flow pattern is captured well by the CF simulation, considering that the CF simulation required about $2$ mins, while the single-porosity simulation took $74$ hrs, a factor of over 2100. 
The average CFL time-step for the CF simulation is about $1.3\ \mathrm{day}$, while the time-steps for the single-porosity simulation are about $7\times10^{-4}\ \mathrm{day}$, which accounts for most of the difference in total CPU time. 
The remaining improvement in CPU time is achieved by the more efficient pressure update, due to the reduced contrast in permeability between the CF and matrix elements. 

\subsection{Example 6: Large 3D domain with ten discrete fractures}\label{sec::ex6}
\begin{figure}
\centerline{\includegraphics[width=.8\textwidth]{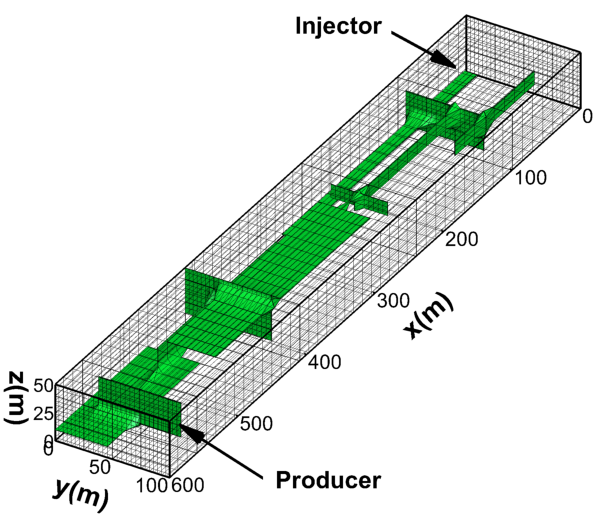}}
\caption{\label{fig::meshex6} Example 6 --- Computational mesh: $600\ \mathrm{m} \times 100\ \mathrm{m} \times 50\ \mathrm{m}$ domain, $1\ \mathrm{m}$ wide CF fracture elements, $48\times 29\times 14$ grid elements. An iso-surface plot of the $50\ \mathrm{d}$ permeability level gives an indication of the locations of the 10 discrete fractures.}
\end{figure}

In this last example, we consider a large $600\ \mathrm{m}\times 100\ \mathrm{m}\times 50\ \mathrm{m}$ three-dimensional domain containing ten discrete planar 2D fractures. Each fracture plane is characterized by two diagonally opposite corners with coordinates $(x_{1}, y_{1}, z_{1})$ and $(x_{2}, y_{2}, z_{2})$, respectively, as provided in Table~\ref{table:: fractures}. We consider $4$ fractures in the $y$-$z$ direction, and $3$ fractures in both the $x$-$y$ and $x$-$z$ orientations. The domain is discretized by $48\times 29\times 14$ elements with $1\ \mathrm{m}$ width of the CF elements. The mesh and locations of the fractures, injection and production wells are illustrated in Figure~\ref{fig::meshex6}. We consider about six orders of magnitude range in spatial scales, and four orders of magnitude in permeability: the fractures have an aperture of $1\ \mathrm{mm}$ and permeability of $100\ \mathrm{d}$, while the matrix blocks have a permeability of $4\ \mathrm{md}$ and porosity of $44\%$. 

\begin{figure*}
\centerline{
\subfigure[$1\%$ PVI]{\includegraphics[width=.5\textwidth]{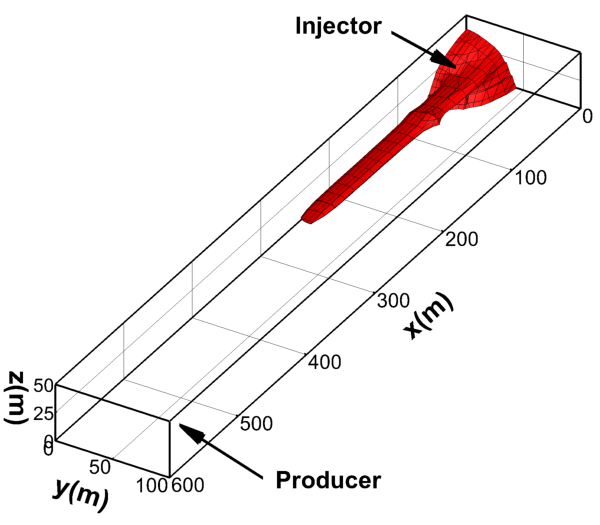}}
\subfigure[$8\%$ PVI]{\includegraphics[width=.5\textwidth]{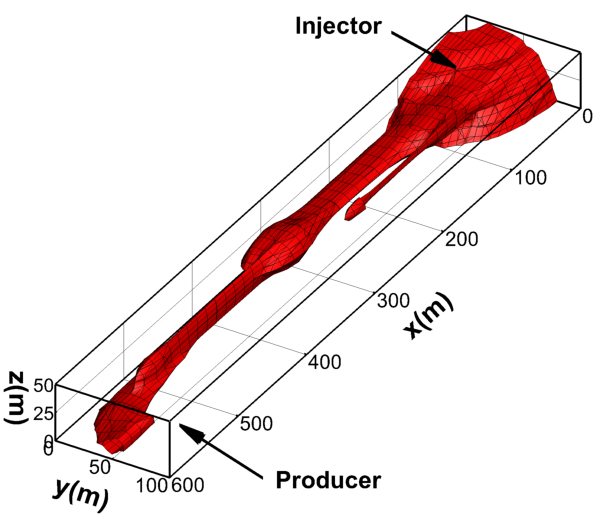}}
}
\centerline{
\subfigure[$25\%$ PVI]{\includegraphics[width=.5\textwidth]{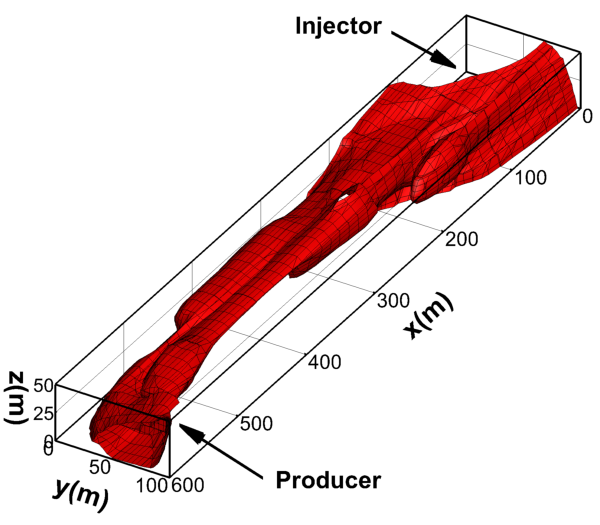}}
\subfigure[$100\%$ PVI]{\includegraphics[width=.5\textwidth]{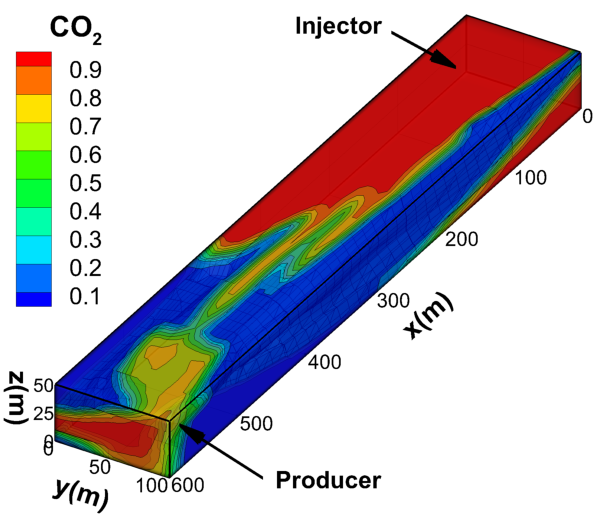}}
}
\caption{\label{fig::CO2compsex6} Example 6 --- Iso-surface plots of the $25$ mol\% level for the overall CO$_{2}$ concentration at $1\%$, $8\%$ and $25\%$ PVI, and contour plot for the overall CO$_{2}$ concentration at $100\%$ PVI.}
\end{figure*}

The domain is initially saturated with the fluid characterized in Table~\ref{table::compdarvishlumped} at a temperature of $400\ \mathrm{K}$ and a pressure of $300\ \mathrm{bar}$ at the bottom. 
In the simulation, one pore volume of CO$_{2}$ is injected at a rate of 10\% PV/yr from the corner at $0\ \mathrm{m}\times 0\ \mathrm{m}\times 0\ \mathrm{m}$ and production is from the opposite corner at $600\ \mathrm{m}\times 100\ \mathrm{m}\times 50\ \mathrm{m}$ at constant pressure. Figure~\ref{fig::CO2compsex6} shows the overall CO$_{2}$ composition throughout the domain at $1\%$, $8\%$, $25\%$ and $100\%$ PVI. Not surprisingly, we find that once the CO$_{2}$ front reaches the first fracture, CO$_{2}$ quickly flows through the connected fractures, resulting in early breakthrough. In this example, we neglected Fickian diffusion, so there is no efficient mechanism for cross-flow between the fractures and the neighboring matrix blocks. 
This example demonstrates that we can apply our discrete fracture model to model compositional multiphase flow in complex three-dimensional discretely fractured domains.

\section{Summary and conclusions}
First, we briefly reiterate the advantages of our higher-order finite element modeling of three-phase flow, and the discrete fracture model. 
The main motivation for the use of the combined MHFE and DG methods is flow in heterogeneous and fractured porous media. From MHFE, we obtain an accurate pressure field across interfaces of difference permeabilities (fractures), because the pressures are continuous across element edges as well as inside the elements through the use of the Raviart-Thomas basis vector fields. At the same time, a continuous velocity field is provided at every point, and to the same order of convergence. This is a marked improvement over lowest-order FD models that only update average  pressures in each element, and compute velocities as a post-process. Phase properties, on the other hand, are intrinsically discontinuous across permeability jumps. The DG method is therefore particularly suitable for the mass transport update, rather than approximating discontinuous properties with continuous methods, even at higher order. Specifically, at edges between a fracture and a matrix block, we have a single continuous pressure and corresponding flux, but a different composition in the fracture from the matrix. Within each element, we use a higher-order approximation to the mass-transport update, which reduces the numerical dispersion as compared to lower-order methods. The higher-order convergence means that accurate results can be obtained on coarser grids and at lower CPU cost.

We adopt the CFE approximation to model fractures, which has considerable advantages over commonly used single- and dual-porosity models. Compared to single-porosity models, the CPU cost is reduced by orders of magnitude by combining fractures with a small amount of matrix fluid into larger computational elements, which relaxes the CFL condition. The CPU time is further improved because the CF elements have a lower contrast in (effective) permeability with the matrix blocks, which speeds up the matrix inversion in the pressure update. As long as the pore volume of the matrix-slice included in the CF elements is small compared to the total pore volume of matrix blocks, results from the CF approximation are indistinguishable from fine mesh simulations. Compared to dual-porosity models, the CF approach has the advantage that there is no need for transfer functions, which may not have a solid physical footing. Interactions between fracture and matrix elements are computed as in single-porosity simulations. Challenging examples that benefit from this approach, and often require discretized matrix blocks, include gravitational and viscous fingering in matrix blocks, gravitational re-infiltration of oil from fractures to neighboring matrix blocks, and Fickian diffusion. The sugar-cube configuration, required by dual-porosity models, may also be overly restrictive, for example in modeling discrete fractures. 

In this work, we have advanced these methods in a number of areas:
\begin{itemize}
\item We have presented the first discrete fracture simulator for fully compositional EOS-based three-phase flow, including both three hydrocarbon phases with transfer of all species between the three phase, and two hydrocarbon phases and an aqueous phase with $\mathrm{CO}_{2}$ solubility in water.
\item We have adopted a CFL condition for compositional multi-phase flow in a fractional flow formulation. The corresponding time-step selection greatly improves the numerical stability of the method compared to earlier work, and in some cases increases the CPU efficiency.
\item Fickian diffusion is modeled with a self-consistent model based on a full matrix of composition dependent diffusion coefficients, derived from irreversible thermodynamics. The open-space diffusion coefficients are reduced by the formation porosity (and tortuosity) to account for the reduced area available for diffusion. Diffusion from fracture to matrix elements is improved, with respect to single-porosity models, by mixing the fracture fluid with a small amount of matrix fluid, which allows fracture-matrix diffusion within the oil and gas phases. Further improvements can be made by enforcing thermodynamic equilibrium \textit{across} the fracture-matrix interface, which we consider in a future work. The generalization of this work to account for capillarity is presented in \citet{moortgatcapillarity}.
\item The detailed description of the MHFE implementation of the CF discrete fracture model includes terms omitted in earlier work, in particular related to the gravitational flux in the fractures. This improves the results when gravity is an important driving force, such as during depletion. The formulation is also significantly simplified by introducing a weighted effective mobility for the CF elements that takes into account both the fracture and the matrix contributions to the flux. In terms of this mobility, the implementation on both unfractured and fractured domains is nearly identical. Our main challenge in modeling fractured domains, particularly in 3D, has been to develop an appropriate mesh generator that also initializes the simulations by working out all the fracture orientations and the intersection areas of fractures with the faces of CF elements. Once these geometric factors are initialized, the modeling of fractured reservoirs is surprisingly straightforward with our proposed model. 
\end{itemize}

\section{Aknowledgements}
We thank the member companies of the Reservoir Engineering Research Institute (RERI) for their support of the work presented in this paper. 

%\bibliography{paper}

\appendix
\section{Coefficients in MHFE expansion of convective fluxes}\label{app::mhfe}
The $\beta_{K, E, E^\prime}$ coefficients in \eqref{eq::qmatrix} for matrix elements are given (as in \citet{moortgatI}) by:
\begin{equation}\label{eq::beta}
\beta_{K, E, E^\prime}=  \lambda_{t,K}\left[\int_{K} \vec{w}_{K,E} \mathrm{K}_{m,K}^{-1} \vec{w}_{K,E^{\prime}} \right]^{-1}.
\end{equation}
\eqref{eq::beta} expresses that for each edge/face $E$, $\beta_{K, E, E}$ is the effective mobility times the length/surface (2D/3D) of edge/face $E$ divided by the distance from the mid-point of $E$ to the center of $K$ (denoted by $L_{E\perp}$).
As an example: on a 2D rectangular mesh, with scalar absolute permeability $\mathrm{K}_m$, and after performing mass-lumping onto the diagonal, \eqref{eq::beta} reduces to:
\begin{equation}\label{eq::betarect}
\beta_{K, E, E^\prime}=2 \mathrm{k}_{t,m,K} \left( \frac{l_y}{l_x}\mathrm{\textbf{I}}_h +  \frac{l_x}{l_y} \mathrm{\textbf{I}}_v \right),
\end{equation}
where we have defined the diagonal matrixes $\mathrm{\textbf{I}}_h = \mathrm{diag}[1,1,0,0]$, $\mathrm{\textbf{I}}_v = \mathrm{diag}[0,0,1,1]$ and $l_{x}$ and $l_{y}$ are the lengths of a rectangular element in the $x$ and $y$ directions, respectively (with edges numbered right $=1$, left $=2$, top $=3$, bottom $ =4$).  

Similarly, for horizontal and vertical fractures (with $\varepsilon$ the area of the fracture intersection with $E$), we have in 2D: 
\begin{equation}\label{eq::betaf}
\beta_{f, e, e^\prime}(\mathrm{hor}) =
2 \varepsilon  \mathrm{k}_{f,K}\mathrm{\textbf{I}}_h/l_x,\quad
\beta_{f, e, e^\prime}(\mathrm{vert}) = 
2 \varepsilon \mathrm{k}_{f,K}\mathrm{\textbf{I}}_v/l_y.
\end{equation}
For a cross-flow element we sum the contributions from the matrix ($\beta_{k, E, E^\prime}$) and fracture ($\beta_{f, e, e^\prime}$) portions. For example, a 2D element $K$ containing a horizontal fracture has
\begin{equation}
\beta_{K, 1, 1} = \beta_{k, 1,1} + \beta_{f, 1,1} =  \mathrm{k}_{t,m}  \frac{l_y-\varepsilon}{l_x/2}\mathrm{\textbf{I}}_h + \frac{\varepsilon}{l_x/2}  \mathrm{k}_{t,f}\mathrm{\textbf{I}}_h, \quad\mbox{and}\quad \beta_{K, 2, 2} = \beta_{K, 1, 1}.
\end{equation}
In terms of the total weighted effective mobility across edge $E$, as defined in \eqref{eq::keff}, we can succinctly write for the mass-lumped $\beta_{K, E, E^{\prime}}$
\begin{equation}
\beta_{K, E, E^{\prime}} = \beta_{K, E, E^{\prime}} \delta_{E,E^{\prime}} = \beta_{k, E, E^\prime} + \beta_{f, e, e^\prime} = \mathrm{k}^{\mathrm{eff}}_{t} \frac{A_{E}}{L_{E\perp}}
\end{equation}
in both 2D and 3D, and for both matrix elements and fracture-containing CF elements. In terms of $\mathrm{k}_{\mathrm{eff}} $, the remaining coefficients in \eqref{eq::qfrac} reduce to the same form as in \eqref{eq::qmatrix}:
\begin{subequations}
\begin{eqnarray}
 \theta_{K, E} &=&\theta_{k, E} + \theta_{f, e}= \sum_{E^\prime}  \beta_{K, E, E^\prime}, \\
 \gamma_{K, E} &=&  \gamma_{k, E} +  \gamma_{f, e} = 
  - \sum_\alpha  \rho_{\alpha,K} \mathrm{k}^{\mathrm{eff}}_{\alpha, K} (\vec{g}\cdot \vec{n}_E) A_{E}.
\end{eqnarray}
\end{subequations}

\section{Coefficients in MHFE expansion of the pressure equation}\label{app::mhfepress}
The coefficients in \eqref{eq::mhfep1} are defined in terms of the weighted effective fractional flow functions $f^{\mathrm{eff}}_{\alpha,K}$ as:
\begin{equation}\label{eq::mif}
m_{i,K} =  \sum_\alpha c_{\alpha, K} x_{i\alpha, K} f^{\mathrm{eff}}_{\alpha,K},\quad\mbox{and}\quad
\vec{s}_{i,K} =  \sum_\alpha c_{\alpha, K}  x_{i\alpha,K} f^{\mathrm{eff}}_{\alpha,K} \vec{G}_{\alpha,K}.
\end{equation}
and in \eqref{eq::mhfep2}, we have
\begin{equation}
\tilde{s}_{i, K, E} = \int_E \vec{s}_{i, K} \cdot \vec{n}_{K, E}.
\end{equation}
$m_{i, K, E}$ is as in \eqref{eq::mif}, but may be evaluated at the edges from the DG results.
We define 
\begin{equation}
\tilde{v}_{K, E} = \sum_{i=1}^{n_c} \bar{v}_{i,K} m_{i, K, E},
\end{equation}
and write the coefficients in \eqref{eq::pdiscr} as:
\begin{subequations}
\begin{eqnarray}
\tilde{\alpha}_K &=& \sum_{E\in K} \tilde{v}_{K, E} \theta_{K, E} , \\
\tilde{\beta}_{K, E} &=& \sum_{E^\prime \in\partial K} \tilde{v}_{K, E}  \beta_{K, E, E^\prime},\\
\tilde{\gamma}_K &=& \sum_{E\in\partial K}\left( \tilde{v}_{K, E} \gamma_{K, E} + \sum_{i=1}^{n_c} \bar{v}_{i,K} \tilde{s}_{i, K, E} \right).
\end{eqnarray}
\end{subequations}

\section{Matrices in global MHFE velocity and pressure systems}\label{app::global}
The matrices $\mathrm{R}$ and $\mathrm{M}$, and vector $\vec{I}$ in \eqref{eq::sysq} collect the coefficients defined in Appendix~\ref{app::mhfe} for each element and edge:
\begin{subequations}\begin{eqnarray}
\mathrm{R} \in \mathbb{R}^{N_K, N_E} &,& R_{K,E} = \theta_{K,E} , \\
\mathrm{M} \in \mathbb{R}^{N_E, N_E} &,& M_{E,E^\prime} = \sum_{K:E,E^\prime\in\partial K} \beta_{K,E,E^\prime},\\
 \vec{I}  \in \mathbb{R}^{N_E} &,& I_E =  \sum_{K:E\in\partial K} \gamma_{K,E} .
\end{eqnarray}\end{subequations}
The matrices in the global system for the pressures, \eqref{eq::sysp}, are defined similarly from the coefficients in Appendix~\ref{app::mhfepress}:
\begin{subequations}\begin{eqnarray}
\mathrm{D} \in \mathbb{R}^{N_K, N_K} &,& D_{KK} = \frac{\zeta}{\Delta t} + \tilde{\alpha}_K,\\
\tilde{\mathrm{R}} \in \mathbb{R}^{N_K, N_E} &,& \tilde{R}_{K,E} = \tilde{\beta}_{K, E}  \\
\vec{G} \in \mathbb{R}^{N_K} &,& G_K = \frac{\zeta p_K (t_\mathrm{old} )}{\Delta t} +  \tilde{\gamma}_K +  
 \sum_i \bar{\nu}_i\left( F_i - \sum_E  \sum_\alpha q^\mathrm{diff}_{i,\alpha, K, E} \right)
\end{eqnarray}\end{subequations}

\newpage

%%%%%%%%%%%% TABLES %%%%%%%%%%%%%%%%%
 \begin{table*}[htdp]
 \caption{\label{table::compex1}Initial composition (mole fraction) $z_i^0$, acentric factor $\omega$, critical temperature $T_c$, critical pressure $p_c$, molar weight $M_w$, critical volume $V_c $ and volume translation $s$ for the fluid characterization in Example 1.}
 \begin{center}
 \begin{tabular*}{\hsize}{@{\extracolsep{\fill}}lrrrrrrr}\hline
 Species &  $z_i^0$ & $\omega$ &$T_c (\mathrm{K})$ & $p_c (\mathrm{bar})$  & $M_w (\mathrm{g}/\mathrm{mole})$& $V_c \left(\mathrm{cm}^3/\mathrm{g}\right)$ & $s$\\\hline
$\mathrm{H}_2\mathrm{O}$				&   0.00	&   0.344	&         647	&         221	&          18	&    2.14	&   0.000\\
$\mathrm{CO}_2$						&   0.00	&   0.239	&         304	&          74	&          44	&    2.14	&   0.020\\
$\mathrm{C}_1\! -\! \mathrm{N}_1$		&   0.45	&   0.011	&         190	&          46	&          16	&    6.14	&  -0.154\\
$\mathrm{C}_2\! -\! \mathrm{C}_3$		&   0.12	&   0.118	&         328	&          47	&          35	&    4.73	&  -0.095\\
$\mathrm{C}_4\! -\! \mathrm{C}_6$		&   0.07	&   0.234	&         458	&          34	&          70	&    4.32	&  -0.047\\
$\mathrm{C}_6\! -\! \mathrm{C}_9$		&   0.08	&   0.370	&         566	&          26	&         108	&    4.24	&   0.038\\
$\mathrm{C}_{10}\! -\! \mathrm{C}_{15}$		&   0.12	&   0.595	&         651	&          19	&         166	&    4.31	&   0.115\\
$\mathrm{CO}_{16+}$						&   0.16	&   1.427	&         824	&          10	&         386	&    3.75	&   0.277\\
 \hline
 \end{tabular*}
 \end{center}
 \end{table*}

 \begin{table*}[htdp]
 \caption{\label{table::densviscex1}Density and viscosity for water, oil and $\mathrm{CO}_{2}$ at $p=300\ \mathrm{bar}$ and $T=350\ \mathrm{K}$ in Example 1.}
  \begin{center}
 \begin{tabular*}{\hsize}{@{\extracolsep{\fill}}l|rrr}\hline
 &	$\mathrm{H}_2\mathrm{O}$  & Oil & $\mathrm{CO}_{2}$  \\\hline
	Density ($\mathrm{g}/\mathrm{cm}^3$)					& 0.985	& 0.713	& 0.754	\\
 Viscosity (cp) 	& 0.36	& 0.53	& 0.03	 \\
 \hline
 \end{tabular*}
 \end{center}
 \end{table*}
 
  \begin{table*}[htdp]
 \caption{\label{table::relperms}Relative permeability parameters for all numerical examples.}
  \begin{center}
 \begin{tabular*}{\hsize}{@{\extracolsep{\fill}}l|rrrrrrrrrrrr}\hline
  & $S_{\mathrm{rw}}^{0}$ &$S_{\mathrm{row}}^{0}$ &$S_{\mathrm{rog}}^{0}$&$S_{\mathrm{rg}}^{0}$ &$k_{\mathrm{rw}}^{0}$& $k_{\mathrm{row}}^{0}$& $k_{\mathrm{rog}}^{0}$& $k_{\mathrm{rg}}^{0}$&$n_{\mathrm{w}}$& $n_{\mathrm{ow}}$&$n_{\mathrm{og}}$& $n_{\mathrm{g}}$\\\hline
 Ex.~1 &  0.0 & 0.5 & 0.1 & 0.0 & 0.3 & 1.0 & 0.6 & 1.0 & 3.0 & 3.0 & 3.5 & 2.4\\
 Exs.~2-3 &  0.3 & 0.5 & 0.1 & 0.0 & 0.3 & 1.0 & 0.4 & 0.6 & 3.0 & 2.0 & 2.0 & 2.0\\
 Exs.~4-5 &  0.05 & 0.2 & 0.2 & 0.05 & 0.65 & 0.5 & 0.5 & 0.65 & 3.0 & 3.0 & 3.0 & 3.0\\
 Ex.~6 & 0 & 0 & 0 & 0 & 1 & 1 & 1  & 1 & 1 & 1 & 1 & 1\\
 \hline
 \end{tabular*}
 \end{center}
 \end{table*}

 \begin{table*}[htdp]
 \caption{\label{table::fluidtor}Initial composition (mole fraction) $z_i^0$, acentric factor $\omega$, critical temperature $T_c$, critical pressure $p_c$, molar weight $M_w$, critical volume $V_c $ and volume translation $s$ for fluid characterization in Example 2.}
 \begin{center}
 \begin{tabular*}{\hsize}{@{\extracolsep{\fill}}l|rrrrrrr}\hline
 Species &  $z_i^0$ & $\omega$ &$T_c (\mathrm{K})$ & $p_c (\mathrm{bar})$  & $M_w (\mathrm{g}/\mathrm{mole})$& $V_c \left(\mathrm{cm}^3/\mathrm{g}\right)$ & $s$\\\hline
$\mathrm{H}_2\mathrm{O}$			&	   0.00	&   0.344	&         647	&         221	&          18	&    2.14	&   0.000\\
$\mathrm{CO}_2$					&	   0.00	&   0.239	&         304	&          74	&          44	&    2.14	&   0.100\\
$\mathrm{C}_1\! -\! \mathrm{N}_1$		&   0.57	&   0.012	&         189	&          46	&          16	&    6.09	&  -0.157\\
$\mathrm{C}_2\! -\! \mathrm{C}_3$	&	   0.16	&   0.120	&         330	&          46	&          35	&    4.73	&  -0.094\\
$\mathrm{C}_4\! -\! \mathrm{C}_6$	&	   0.08	&   0.233	&         455	&          35	&          69	&    4.32	&  -0.048\\
$\mathrm{C}_6\! -\! \mathrm{C}_{10}$		&   0.09	&   0.428	&         584	&          24	&         120	&    4.25	&   0.055\\
$\mathrm{CO}_{11+}$					&	   0.11	&   1.062	&         751	&          13	&         293	&    4.10	&   0.130\\
 \hline
 \end{tabular*}
 \end{center}
 \end{table*}

 \begin{table*}[htdp]
 \caption{\label{table::densvisc300bar}Density and viscosity for water, oil and $\mathrm{CO}_{2}$ at $p=300$ and $350\ \mathrm{bar}$ and $T=350\ \mathrm{K}$ in Example 2 and at $p=350\ \mathrm{bar}$ and $T=400\ \mathrm{K}$ in Example 3.}
  \begin{center}
 \begin{tabular*}{\hsize}{@{\extracolsep{\fill}}ll|rrrr}\hline
 &	& $\mathrm{H}_2\mathrm{O}$ & Oil  & $\mathrm{CO}_{2}$ & Gas  \\\hline
Density ($\mathrm{g}/\mathrm{cm}^3$),& $p=350\ \mathrm{bar}$, $T=350\ \mathrm{K}$					& 0.987	& 0.586	& 0.841 & --	\\
 	Viscosity (cp),& $p=350\ \mathrm{bar}$, $T=350\ \mathrm{K}$ 	& 0.37	& 0.20	& 0.04 & --	 \\\hline
Density ($\mathrm{g}/\mathrm{cm}^3$),& $p=300\ \mathrm{bar}$, $T=350\ \mathrm{K}$					& 0.985	& 0.602	& 0.782	& 0.260 \\
	Viscosity (cp),&$p=300\ \mathrm{bar}$, $T=350\ \mathrm{K}$ 		& 0.37	& 0.23	& 0.03	& 0.03 \\\hline
Density ($\mathrm{g}/\mathrm{cm}^3$),& $p=300\ \mathrm{bar}$, $T=400\ \mathrm{K}$ 		& 0.953	& 0.567	& 0.560	& 0.237 \\
	Viscosity (cp),&$p=300\ \mathrm{bar}$, $T=400\ \mathrm{K}$ 		& 0.22	& 0.16	& 0.03	& 0.03 \\
 \hline
 \end{tabular*}
 \end{center}
 \end{table*}
 
    \begin{table*}[htdp]
 \caption{\label{table:: fractures}Fracture characterization in Example 6.}
 \begin{center}
 \begin{tabular*}{\hsize}{@{\extracolsep{\fill}}lrrrrrr}\hline
  Index &  $x_{1}$ (m) &$y_{1}$ (m) & $z_{1}$ (m)  & $x_{2}$(m)&  $y_{2}$ (m) & $z_{2}$ (m)\\\hline
1 & 100 & 5 & 5 &   100 & 80 & 45  \\
2 & 240 & 25 & 25  & 240 & 75 & 40  \\
3 & 410 & 0 & 10  & 410 & 80 & 50  \\
4 & 560 & 10 & 5 & 560 & 90 & 45 \\
5 & 300 & 25 & 5 & 500 & 25 & 35 \\
6 & 100 & 50 & 25 & 250 & 50 & 45 \\
7 & 10 & 75 & 25 & 150 & 75 & 45 \\
8 & 450 & 0 & 10 & 600 & 50 & 10 \\
9 & 10 & 10 & 20 & 300 & 30 & 20 \\
10 & 250 & 30 & 40 & 500 & 80 & 40\\
 \hline
 \end{tabular*}
 \end{center}
 \end{table*}
 
    \begin{table*}[htdp]
 \caption{\label{table::compdarvishlumped}Initial composition (mole fraction) $z_i^0$, acentric factor $\omega$, critical temperature $T_c$, critical pressure $p_c$, molar weight $M_w$, and volume translation $s$ for fluid characterization in Example 6 (we note that the volume shift for $\mathrm{C}_{35+}$ may be lower than for a real reservoir oil).}
 \begin{center}
 \begin{tabular*}{\hsize}{@{\extracolsep{\fill}}l|rrrrrr}\hline
 Species &  $z_i^0$  & $\omega$ &$T_c (\mathrm{K})$ & $p_c (\mathrm{bar})$  & $M_w (\mathrm{g}/\mathrm{mole})$&  $s$\\\hline
$\mathrm{CO}_2$						& 0.0083	&   0.239	&         304.	&          74.	&          44.&      0.020\\
$\mathrm{C}_1\! -\! \mathrm{N}_2$		& 0.4427	&   0.011	&         190.	&          46.	&          16.	&     -0.154\\
$\mathrm{C}_2\! -\! \mathrm{C}_3$		& 0.1177	&   0.118	&         328.	&          47.	&          35.	&     -0.095\\
$\mathrm{C}_4\! -\! \mathrm{C}_6$		& 0.0741	&   0.234	&         458.	&          34.	&          70.	&     -0.047\\
$\mathrm{C}_7\! -\! \mathrm{C}_9$		& 0.0821	&   0.370	&         566.	&          26.	&         108.	&      0.038\\
$\mathrm{C}_{10}\! -\! \mathrm{C}_{15}$		& 0.1158	&   0.595	&         651.	&          19.	&         166.	&      0.115\\
$\mathrm{C}_{16}\! -\! \mathrm{C}_{22}$		& 0.0551	&   0.870	&         717.	&          16.	&         247.	&      0.169\\
$\mathrm{C}_{23}\! -\! \mathrm{C}_{34}$		& 0.0465	&   1.060	&         797.	&          14.	&         336.	&      0.223\\
$\mathrm{C}_{35+}$		& 0.0577	&   1.100	&         958.	&          10.	&         558.	&      0.010\\
 \hline
 \end{tabular*}
 \end{center}
 \end{table*}

%%%%%%%%%%%% FIGURES %%%%%%%%%%%%%%%%%

\clearpage


\end{document}